\documentclass[letter]{article}[11pt]

\usepackage{amssymb,amsfonts,amsmath,amsthm,amscd,dsfont,mathrsfs}
\usepackage{eqsection,graphicx,float,psfrag,epsfig,color}
\usepackage{hyperref}

\newtheorem{claim}{Claim}
\newtheorem{lemma}{Lemma}

\newtheorem{theorem}{Theorem}
\newtheorem{proposition}[claim]{Proposition}
\newtheorem{corollary}[claim]{Corollary}
\newtheorem{definition}[claim]{Definition}

\newtheorem{remark}{Remark}

\definecolor{Red}{rgb}{1,0,0}
\definecolor{Blue}{rgb}{0,0,1}
\definecolor{Olive}{rgb}{0.41,0.55,0.13}
\definecolor{Green}{rgb}{0,1,0}
\definecolor{MGreen}{rgb}{0,0.8,0}
\definecolor{DGreen}{rgb}{0,0.55,0}
\definecolor{Yellow}{rgb}{1,1,0}
\definecolor{Cyan}{rgb}{0,1,1}
\definecolor{Magenta}{rgb}{1,0,1}
\definecolor{Orange}{rgb}{1,.5,0}
\definecolor{Violet}{rgb}{.5,0,.5}
\definecolor{Purple}{rgb}{.75,0,.25}
\definecolor{Brown}{rgb}{.75,.5,.25}
\definecolor{Grey}{rgb}{.5,.5,.5}
\definecolor{Pink}{rgb}{1,0,1}
\definecolor{DBrown}{rgb}{.5,.34,.16}

\definecolor{Black}{rgb}{0,0,0}

\newcommand{\enp} {\hfill \rule{2.2mm}{2.6mm}}

\newcommand{\G}{\Gamma}
\newcommand{\g}{\gamma}
\newcommand{\Om}{\Omega}
\newcommand{\om}{\omega}
\newcommand{\al}{\alpha}
\newcommand{\la}{\lambda}

\newcommand{\p}{\partial}
\newcommand{\ep}{\epsilon}
\newcommand{\no}{\nonumber}
\newcommand{\mc}{\mathcal}
\newcommand{\mb}{\mathbb}
\newcommand{\mf}{\mathfrak}
\newcommand{\mbf}{\mathbf}
\newcommand{\mrm}{\mathrm}
\newcommand{\f}{\frac}
\newcommand{\stm}{\setminus}
\def\dv{d_{\mrm{TV}}}
\def\ct{\mf{c}}
\def\dt{\mf{d}}
\def\et{\mf{e}}

\title{A Sequential Algorithm for Generating Random Graphs}
\date{}
\author{Mohsen Bayati\thanks{Stanford University, Stanford, CA 94305; bayati@stanford.edu}
\and Jeong Han Kim\thanks{Yonsei University, South Korea; jehkim@yonsei.ac.kr} \and Amin Saberi
\thanks{Stanford University, Stanford, CA 94305; saberi@stanford.edu.}}
\begin{document}
\maketitle

\begin{abstract}
We present a nearly-linear time algorithm for counting and randomly generating simple graphs
with a given degree sequence in a certain range. For degree sequence
$(d_i)_{i=1}^n$ with maximum degree $d_{\max}=O(m^{1/4-\tau})$, our
algorithm generates almost uniform random graphs with that degree sequence
in time $O(m\,d_{\max})$ where $m=\f{1}{2}\sum_id_i$ is the number of
edges in the graph and $\tau$ is any positive constant.  The fastest
known algorithm for uniform generation of these graphs \cite{MckayWormald1990b} has a running
time of $O(m^2d_{\max}^2)$. Our method also gives an independent proof of
McKay's estimate \cite{Mckay} for the number of such graphs.

%Our approach is based on the \emph{sequential importance sampling} (SIS)
%technique which has been recently successful for counting graphs
%\cite{ChenDiaconisHolmsLiu,JoePersi,blanchet}. Unfortunately
%validity of the SIS method is only known through simulations and our
%work together with \cite{blanchet} are the first results that
%analyze the performance of these methods.

We also use sequential importance sampling to
derive fully Polynomial-time Randomized Approximation Schemes (FPRAS)
for counting and uniformly generating random graphs for the same range of $d_{\max}=O(m^{1/4-\tau})$.

Moreover, we show that for $d = O(n^{1/2-\tau})$, our algorithm can
generate an asymptotically uniform $d$-regular graph. Our results
improve the previous bound of $d = O(n^{1/3-\tau})$ due to Kim and Vu
\cite{KimVuSandwitch} for regular graphs.

\end{abstract}
%\thispagestyle{empty}
%\newpage
%\setcounter{page}{1}
%\vspace{5mm}
\section{Introduction}\label{sec:intro}
The focus of this paper is on generating random simple graphs
(graphs with no multiple edges or self loop) with a given degree
sequence. Random graph generation has been studied extensively as an
interesting theoretical problem (see \cite{Wormald1999,JoePersi}
for detailed surveys). It has also become an important tool in a
variety of real world applications including detecting motifs in
biological networks \cite{MSOIKCA}
and simulating networking protocols on the Internet topology
\cite{inet,faloutsos,medina,bu,doyle}. The best algorithm for this problem was
given by McKay and Wormald \cite{MckayWormald1990b} that uses certain switches on the configuration model and
produces random graphs with uniform distribution in $O(m^2d_{\max}^2)$ time. However, this running time can be slow for the networks with millions of edges.  This has constrained practitioners to use simple heuristics that are non-rigorous and have often led to
wrong conclusions \cite{MiloKashtanItzkovitzNewmanAlon,MSOIKCA}. Our main contribution in
this paper is to provide a nearly-linear time, fully polynomial randomized approximation scheme (FPRAS)
for generating random graphs. An FPRAS provides an arbitrary close approximation in time that depends only polynomially on the input size and the desired error. (For precise definitions of FPRAS, see Definition 1 in Section \ref{sec:algs}.)

Recently, sequential importance sampling (SIS) has been
suggested as a more suitable approach for designing fast graph generation algorithms \cite{ChenDiaconisHolmsLiu,JoePersi,knuth,PersiBassetti}. Chen \emph{et
al.}~\cite{ChenDiaconisHolmsLiu} used the SIS method to generate
bipartite graphs with a given degree sequence. Later Blitzstein and
Diaconis \cite{JoePersi} used a similar approach for generating general
graphs with given degrees. But these results are mostly empirical, and in a few cases SIS is shown to be slow \cite{BezakovaSinclairStephanovicVigoda}. However, the
simplicity of these algorithms and their great performance in several
instances suggest that a further study of the SIS method is necessary.

\paragraph{The Result.}  Let $d_1,\ldots,d_n$ be non-negative integers with $\sum_{i=1}^n d_i = 2m$. Our algorithm for generating a graph with degree sequence $d_1,\ldots,d_n$ is a generalization of Steger and Wormald's algorithm for regular graphs \cite{StegerWormald}. It works as follows: start with an empty graph and sequentially add edges
between the pairs of non-adjacent vertices. In every step of the
procedure, the probability that an edge is added between two
distinct vertices $i$ and $j$ is proportional to $\hat{d}_i
\hat{d}_j (1 - d_i d_j/4m)$ where $\hat{d}_i$ and $\hat{d}_j$ denote
the remaining degrees of vertices $i$ and $j$. The remaining degree of a vertex $i$ is equal to $d_i$ minus its current degree.
We will show that this algorithm produces an asymptotically uniform
sample with running time $O(m\,d_{\max})$ when
$d_{\max}=O(m^{1/4-\tau})$ and $\tau$ is any positive constant. Then, we use SIS to obtain an FPRAS for any $\ep,\delta>0$ with running time
$O(m\,d_{\max}\ep^{-2}\log(1/\delta))$.
{The same result holds when the algorithm is used for generating bipartite graphs.}
Moreover, we show that for $d = O(n^{1/2-\tau})$, this algorithm can
generate $d$-regular graphs with an asymptotically uniform distribution. Our results
improve the bounds of Kim and Vu \cite{KimVuSandwitch} and
Steger and Wormald \cite{StegerWormald} for the regular graphs.

\paragraph{Related Work.} McKay and Wormald \cite{Mckay,MckayWormald1991} give asymptotic estimates for the number of graphs
with $d_{\max}=O(m^{1/3-\tau})$. However, the error terms in their estimates are larger
than what is needed to apply Jerrum, Valiant and Vazirani's
\cite{JerrumValiantVazirani} reduction to achieve an asymptotically uniform sampling. Jerrum and Sinclair \cite{JerrumSinclair1989}, however, use
a random walk on the self-reducibility tree and give an FPRAS for
uniformly sampling the graphs with $d_{\max}=o(m^{1/4})$. The running
time of their algorithm is $O(m^3n^2\ep^{-2}\log(1/\delta))$ \cite{AlistairEmail}. A
different random walk that has been studied by \cite{JerrumSinclair1990,JerrumSinclairMcKay,BezakovaBhatnagarVigoda}, gives an FPRAS for
the random generation of bipartite graphs with all degree sequences and
general graphs with almost all degree sequences. However, the running
time of all these algorithms is at least O($n^4m^3d_{\max}\ep^{-2}\log(1/\delta)$). Other Markov chains methods are also studied in \cite{CooperDyerGreenhill,DiaconisGangolli,GkantsidisMihailZegura}.

McKay and Wormald also introduced an algorithm based on a certain switching technique on the configuration model
that achieves the best performance \cite{MckayWormald1990b}. It produces random graphs with uniform distribution
(better than FPRAS) and has a faster running time. Their algorithm works for graphs with
$d_{\max}^3=O(m^2/\sum_id_i^2)$ and $d_{\max}^3=o(m+\sum_id_i^2)$
with an average running time of $O(m+(\sum_id_i^2)^2)$. This leads to an
$O(n^2d^4)$ average running time for $d$-regular graphs with
$d=O(n^{1/3})$.

Very recently and independently from our work,
Blanchet \cite{blanchet} has used McKay's estimate \cite{Mckay} and SIS
technique to obtain an FPRAS with running time of $O(m^2\ep^{-2}\log(1/\delta))$ for counting
bipartite graphs with given degrees when $d_{\max}=o(m^{1/4})$. His
work is based on defining an appropriate Lyapunov function as well
as using Mckay's estimate.

\paragraph{Our Technical
Contribution.} Our algorithm and its analysis are based on the beautiful
works of Steger and Wormald \cite{StegerWormald} and Kim and Vu
\cite{Kim-Vu}. The technical contributions of our work beyond their
analysis are as follows:
\begin{enumerate}
\item In both \cite{StegerWormald,Kim-Vu} the output
distribution of the proposed algorithms are asymptotically uniform. Here
we use SIS technique to obtain an FPRAS.

\item Both \cite{StegerWormald,Kim-Vu} use McKay's estimate
\cite{Mckay} in their analysis. In this paper we give a
combinatorial argument to control the failure probability of the
algorithm and obtain a new proof for McKay's estimate.

\item We exploit the combinatorial structure and use a martingale tail
inequality to show the concentration results for $d$-regular graphs with
$d=O(n^{1/2-\tau})$ where the previous polynomial
inequalities \cite{Kim-Vu2000} do not work.
\end{enumerate}

\paragraph{Other Applications and Extensions.}  Our algorithm and its
analysis provide more insight into the modern random graph models,
such as the configuration model or the random graphs with a given
\emph{expected} degree sequence \cite{ChungLu}. In these models, the
probability of having an edge between vertices $i$ and $j$ of the
graph is proportional to $d_id_j$. However, one can use our analysis
or McKay's formula \cite{Mckay} to see that in a random simple graph,
this probability is proportional to $d_i d_j (1 - d_i d_j /2m)$. We
expect that by adding the correction term and using the
concentration result of this paper, it is possible to obtain sandwiching theorems similar to
\cite{KimVuSandwitch}.

In a follow up work, Bayati et al. \cite{BMS07} uses similar ideas to generate random graphs with large
girth. These graphs are useful for designing high performance Low-Density Parity-Check (LDPC) codes (cf. \cite{Montanari}).

\paragraph{Organization of the Paper.} The rest of the paper has
the following structure. The algorithm and the main results are
stated in Section \ref{sec:algs}. In Section \ref{sec:idea}, we
explain the intuition behind the weighted configuration model and
our algorithm while also describing the SIS approach.
Finally Sections \ref{sec:analysis}-\ref{sec:Kim-Vu-Calc} are dedicated to the analysis and the proofs.

\section{Our Algorithm}\label{sec:algs}

Suppose that $n$ nonnegative integers $d_1,
d_2,\ldots d_n$ with $\sum_{i=1}^nd_i=2m$ are given.
Assume that this sequence is also \emph{graphical}. That is, there exists at least
one simple graph with these degrees.
We propose the following procedure for sampling (counting) an element (the number of elements)
of the set $\mc{L}(\bar{d})$ of all labeled simple graphs $G$ with vertices $V=\{v_1,v_2,\ldots,v_n\}$
and degree sequence $\bar{d}$ =$(d_1, d_2, \cdots, d_n)$.
Throughout this paper $m=\sum_{i=1}^nd_i/2$ is the number of edges in the graph, $d_{\max}=\max_{i=1}^n\{d_i\}$
and for the regular graphs, $d$ refers to the degrees; i.e. $d_i=d$ for all
$i=1,\ldots,n$.  We denote the set of all $d$-regular graphs with $n$ vertices by $\mc{L}(n,d)$.
\vspace{.1in} \hrule \vspace{.07in}
\noindent Procedure {\bf $\mrm{A}$}
\vspace{.07in} \hrule \vspace{.1in}
\noindent {\tt INPUT:} A graphical degree sequence $\bar{d}=(d_1, d_2, \cdots, d_n)$.\\
\noindent {\tt OUTPUT:} A graph $G$ with degree sequence $\bar{d}$ or failure. An approximation $N$ for the number of graphs with degree sequence $\bar{d}$ or $0$.
\begin{itemize}
\item[(1)] Let $E$ be a set of edges, $\hat{d}=(\hat{d}_1,\ldots,\hat{d}_n)$ be an $n$-tuple of
integers and $P$ be a real number. Initialize them by $E=$ Empty set, $\hat{d}=\bar{d},$ and $P=1$.
\item[(2)] Choose two vertices $v_i,v_j\in V$ with probability proportional to
$\hat{d}_i\hat{d}_j(1-\f{d_id_j}{4m})$ among all pairs $v_i,v_j$ with
$i\neq j$ and $\{v_i,v_j\}\notin E$. Denote this probability by  $p_{ij}$. Multiply $P$ by $p_{ij}$, add $\{v_i,v_j\}$ to $E$ and reduce
each of $\hat{d}_i,~\hat{d}_j$ by $1$.
\item[(3)] Repeat step (2) until no more edges can be added to $E$.
\item[(4)] If $|E|<m$ report \emph{failure} and output $N=0$, otherwise output
$G=(V,E)$ and $N=(m!~P)^{-1}$.
\end{itemize}
\vspace{.1in}
\hrule \vspace{.1in}
Note that for the regular graphs the factors $1-d_id_j/4m$ are redundant
and Procedure $\mrm{A}$ is the same as Steger and Wormald's \cite{StegerWormald} algorithm.
The next two theorems characterize the output distribution of Procedure $\mrm{A}$.
%---------------------------------------------------- MAIN THEOREM
\begin{theorem}\label{thm:main-nobias} For an arbitrary number $\tau>0$ and
for any degree sequence $\bar{d}$ with maximum degree of
$O(m^{1/4-\tau})$, Procedure $\mrm{A}$ can be implemented so that
it terminates successfully with probability $(1-o(1))$ in expected running time O($m\,d_{\max}$).
Furthermore, any graph $G$ with degree
sequence $\bar{d}$ is generated with a probability within $1\pm o(1)$
factor of the uniform probability.
\end{theorem}
For the regular graphs a similar result can be shown in a larger
range for the degrees.
\begin{theorem}\label{thm:main-regular} For an arbitrary number $\tau>0$ and for
$d=O(n^{1/2-\tau})$, Procedure $\mrm{A}$ generates all graphs $G$ in $\mc{L}(n,d)$
with probability within $1\pm o(1)$ factor of the uniform probability,
except for the graphs in a subset of size $o(|\mc{L}(n,d)|)$.
In other words as $n\to\infty$, the output distribution of Procedure
$\mrm{A}$ converges to the uniform distribution in total variation distance.
\end{theorem}
The results above show that the output distribution of Procedure
$\mrm{A}$ is close to uniform only when $n$ is sufficiently large. Nevertheless, it is desirable to be close to uniform for all values of $n$. In order to do that,
we find an FPRAS for calculating $|\mc{L}(\bar{d})|$ and also for randomly generating the elements of $\mc{L}(\bar{d})$.
\begin{definition}\label{def:FPRAS}
An FPRAS for approximately counting graphs with degree sequence $\bar{d}$ is an algorithm
that for any $\ep,\delta>0$, outputs an estimate $X$ for $|\mc{L}(\bar{d})|$ where $\mb{P}\{(1-\ep)|\mc{L}(\bar{d})|\leq
X\leq (1+\ep)|\mc{L}(\bar{d})|\}\geq 1-\delta$, and has a running time polynomial in $m,1/\ep,\log(1/\delta)$.

Similarly, an FPRAS for randomly generating graphs with degree sequence $\bar{d}$ is an algorithm that for any $\ep>0$, has a running time polynomial in $m,1/\ep$, and with probability at least $0.5$, it outputs a graph from the set $\mc{L}(\bar{d})$ with probability within $\frac{1\pm\ep}{c}$ of the uniform where $c$ is a constant.

Throughout this paper we assume $0<\ep,\delta<1$ and for convenience, we define a real valued random variable $X$ to be an
$(\ep,\delta)$-estimate for a number $y$ if~ $\mb{P}\{(1-\ep)y\leq
X\leq (1+\ep)y\}\geq 1-\delta$.
\end{definition}
The following theorem summarizes our main result.
\begin{theorem}\label{thm:fpras} For an arbitrary number $\tau>0$, degree sequence $\bar{d}$ with maximum degree of
$O(m^{1/4-\tau})$, and any $\ep,\delta >0$ the algorithm CountGraphs of Section \ref{sec:idea} is an FPRAS with an expected running time of $O(m\,d_{\max}\ep^{-2}\log(1/\delta))$ for
counting graphs with degree sequence $\bar{d}$.  Moreover, the algorithm GenerateGraph of Section \ref{sec:idea} is an FPRAS with an expected running time of $O(m\,d_{\max}\ep^{-2})$ for randomly generating graphs with degree sequence $\bar{d}$.
\end{theorem}
{
\begin{remark}\label{rem:bipartite}
For generating bipartite graphs, step (2) of Procedure $\mrm{A}$ should be modified to
\begin{itemize}
\item[(2)] Choose two vertices $v_i,v_j\in V$ with probability proportional to
{$\hat{d}_i\hat{d}_j(1-\f{d_id_j}{2m})$} among all pairs $v_i,v_j$ with
$\{v_i,v_j\}\notin E$, and $v_i,v_j$ not belonging to the same part of the graph. Denote this probability by  $p_{ij}$ and
multiply $P$ by $p_{ij}$. Add $\{v_i,v_j\}$ to $E$ and reduce
each of $\hat{d}_i,~\hat{d}_j$ by $1$.
\end{itemize}
Then corresponding versions of Theorems \ref{thm:main-nobias}-\ref{thm:fpras} can be shown.
\end{remark}
}

%------------------------------------------------------------------------------
\section{Definitions and the Main Idea}\label{sec:idea}

Before explaining our approach let us quickly review the
configuration model (cf. \cite{BenderCanfield,Bollobas1980,MolloyReed1995} for more details). Let $W=\cup_{i=1}^n W_i$ be a set of
$2m=\sum_{i=1}^n d_i$ labeled mini-vertices with $|W_i|=d_i$. Consider a procedure
that finds a random perfect matching $\mc{M}$ between mini-vertices by choosing pairs of mini-vertices sequentially and uniformly at random.
Such a matching is also called a \emph{configuration} on $W$. We can see that the number of all distinct configurations is equal
to $(1/m!)\prod_{r=0}^{m-1}{2m-2r\choose 2}$. Given a configuration $\mc{M}$, we can obtain a graph
$G_\mc{M}$ with degree sequence $\bar{d}$ by  combining
the mini-vertices of each $W_i$ to form a vertex $v_i$, .

Note that the graph $G_\mc{M}$ might have self edge loops or
multiple edges. In fact McKay and Wormald's estimate
\cite{MckayWormald1991} shows that this happens with very high
probability except when $d_{\max}=O(\log^{1/2} m)$. In order to fix
this problem, Steger and Wormald \cite{StegerWormald} proposed that at any step one can only look at
those pairs of mini-vertices that lead to simple graphs (denote
these by \emph{suitable pairs}) and pick one uniformly at random.
For $d$-regular graphs when $d=O(n^{1/28-\tau})$ Steger and Wormald
have shown that this approach asymptotically samples
regular graphs with uniform distribution and Kim and Vu \cite{Kim-Vu}
have extended that to $d=O(n^{1/3-\tau})$.

\subsection{Weighted configuration model}

Unfortunately, when the degree sequence is not uniform, the above procedure
may generate some graphs with a probability exponentially larger (or smaller) than uniform probability.
In this paper, we will show
that for non-regular degree sequences suitable pairs should be
picked non-uniformly. In fact, Procedure $\mrm{A}$ is a weighted
configuration model where at any step a suitable pair $\{u,v\}$ with $u\in W_i$ and $v\in W_j$ is picked with probability proportional to
$1-d_id_j/4m$.

Here is a rough intuition behind Procedure $\mrm{A}$. Define the
\emph{execution tree} $\mrm{T}$ of the configuration model as
follows: Consider a rooted tree where its root (the vertex at level
zero) corresponds to the empty matching in the beginning of the
model and level $r$ vertices correspond to all partial matchings
that can be constructed after $r$ steps. There is an edge in
$\mrm{T}$ between a partial matching $\mc{M}_r$ from level $r$ to a
partial matching $\mc{M}_{r+1}$ from level $r+1$ if
$\mc{M}_r\subset\mc{M}_{r+1}$. Any path from the root to a leaf of
$\mrm{T}$ corresponds to one possible way of generating a random
configuration.

Let us denote those partial matchings $\mc{M}_r$ { whose} corresponding
partial graph $G_{\mc{M}_r}$ is simple by ``valid'' matchings and denote the remaining
partial matchings by ``invalid''.
Our goal is to sample valid leaves of the tree
$\mrm{T}$ uniformly at random. Steger and Wormald's improvement to
the configuration model is to restrict the algorithm at step $r$ to the
valid children of $\mc{M}_r$ and picking one uniformly at random.
This approach leads to an almost uniform generation for the regular graphs
\cite{StegerWormald,Kim-Vu} since the number of valid children for
all partial matchings at level $r$ of $\mrm{T}$, is almost
equal. However, it is crucial to note that for non-regular degree
sequences if the $(r+1)^{th}$-edge matches two elements belonging to the
vertices with larger degrees, the number of valid children for
$\mc{M}_{r+1}$ will be smaller. Thus, there will be a bias towards
graphs that have more of such edges.

In order to find a rough estimate of the bias, fix a graph $G$ with
degree sequence $\bar{d}$.  Let $\mrm{M}(G)$ be the set of all leaves
$\mc{M}$ of the tree $\mrm{T}$ that lead to graph $G$; i.e.
those configurations $\mc{M}$ with $G_{\mc{M}}=G$. It is easy to see that
$|\mrm{M}(G)|=m!\prod_{i=1}^nd_i!$. Moreover, for exactly
$(1-q_r)\,|\mrm{M}(G)|$ of these leaves, a fixed edge $\{i,j\}$ of $G$
appears in the first $r$ steps of the path leading to them; i.e.
$\{i,j\}\in\mc{M}_r$.  Here $q_r=(m-r)/m$.  Furthermore, we can show
that for a typical matching after step $r$, the number of unmatched
mini-vertices in each $W_i$ is roughly $d_iq_r$. Thus the expected
number of unsuitable pairs $\{u,v\}$ is about
$\sum_{i\sim_Gj}d_id_jq_r^2(1-q_r)$. Similarly, the expected number of
unsuitable pairs corresponding to self edge loops is approximately
$\sum_{i=1}^n{d_iq_r\choose2}\approx 2mq_r^2\la(\bar{d})$ where
$\la(\bar{d})=\sum_{i=1}^n{d_i\choose2}/(\sum_{i=1}^nd_i)$.
Therefore, defining $\gamma_G=\sum_{i \sim_G j}d_id_j/4m$ and using
${2m-2r\choose2}\approx 2m^2q_r^2$ we can approximate $\mb{P}_{\mrm{A}}(G)$, the probability of generating $G$ with Procedure $\mrm{A}$ by
\begin{eqnarray*}
\mb{P}_{\mrm{A}}(G)&\approx&m!\left(\prod_{i=1}^nd_i!\right)
\prod_{r=0}^{m-1}\f{1}{2m^2q_r^2-2mq_r^2\la(\bar{d})-4m(1-q_r)q_r^2\gamma_G}\\
&\approx& e^{\la(\bar{d})+\gamma_G}~m!\left(\prod_{i=1}^nd_i!\right)
\prod_{r=0}^{m-1}\f{1}{{2m-2r\choose2}}~\propto~~ e^{\gamma_G}.
\end{eqnarray*}
Hence, adding the edge $\{i,j\}$ roughly creates an $\exp(d_id_j/4m)$
bias. To cancel that effect we need to reduce the probability of picking
$\{i,j\}$ by $\exp(-d_id_j/4m)\approx1-d_id_j/4m$. We will rigorously
prove the above argument in Section \ref{sec:analysis}. %\vspace{5mm}

\subsection{Obtaining a fully polynomial randomized approximation scheme}

The output distribution of Procedure $\mrm{A}$ denoted by $\mb{P}_{\mrm{A}}$ is asymptotically uniform. But when $m$ is small, it is desirable to reduce the deviation of the output distribution from the uniform distribution. Note that it is not possible to use an accept/reject scheme to obtain uniform distribution since the probability $\mb{P}_{\mrm{A}}(G)$ is not known for any given graph $G$. In fact, for an output $G$ of Procedure $\mrm{A}$, the variable $P$ is the probability of generating \emph{one ordering} of the edges of $G$ among all $m!$ possible permutations. Different orderings can have probabilities that vary exponentially which further complicates the calculation of $\mb{P}_{\mrm{A}}(G)$.

However, we can use the Sequential Importance Sampling (SIS) method, similar to \cite{ChenDiaconisHolmsLiu}, to find very close estimates for $\mb{P}_{\mrm{A}}(G)$ and
$|\mc{L}(\bar{d})|$. Then with a simple accept/reject scheme we can obtain a distribution that is very close to the uniform distribution. For example if $\mb{P}_{\mrm{A}}(G)|\mc{L}(\bar{d})|\geq 1$ then we can accept graph $G$ with probability $\left(\mb{P}_{\mrm{A}}(G)|\mc{L}(\bar{d})|\right)^{-1}$. This approach will be explained in more detail in this section.

\subsubsection{FPRAS for Counting via SIS.}

Denote the set of all orderings $\mc{N}$ that lead to a graph in $\mc{L}(\bar{d})$ by $\mc{K}(\bar{d})$.
Therefore, $|\mc{K}(\bar{d})|=m!\,|\mc{L}(\bar{d})|$.
Let $\mb{Q}$ be the uniform distribution on $|\mc{K}(\bar{d})|$. Procedure $\mrm{A}$ samples an ordering  $\mc{N}\in \mc{K}(\bar{d})$ from a ``trial distribution'' $\mb{P}_{\mrm{A}}$, where $\mb{P}_{\mrm{A}}(\mc{N})>0$ for all $\mc{N}\in \mc{K}(\bar{d})$.
Thus, we have
\begin{eqnarray*}
\mb{E}_{\mb{P}_{\mrm{A}}}(\frac{1}{\mb{P}_{\mrm{A}}})=\sum_{\mc{N}\in \mc{K}(\bar{d})}\frac{1}{\mb{P}_{\mrm{A}}(\mc{N})}\mb{P}_{\mrm{A}}(\mc{N})=|\mc{K}(\bar{d})|.
\end{eqnarray*}
Hence, we can estimate $|\mc{K}(\bar{d})|$ by
$$\widehat{|\mc{K}(\bar{d})|}=\frac{1}{k}\sum_{i=1}^k\frac{1}{\mb{P}_{\mrm{A}}(\mc{N}_i)}$$
from $k$ iid samples $\mc{N}_1,\ldots,\mc{N}_k$ drawn from $\mb{P}_{\mrm{A}}(\mc{N})$. Now in order to estimate $|\mc{L}(\bar{d})|=|\mc{K}(\bar{d})|/m!$ we can use
$$\widehat{|\mc{L}(\bar{d})|}=\frac{1}{k}\sum_{i=1}^k\frac{1}{m!\mb{P}_{\mrm{A}}(\mc{N}_i)}.$$
Note that when an ordering $\mc{N}$ is the output of Procedure $\mrm{A}$ then the number $N$,
that is also an output of Procedure $\mrm{A}$, is equal to $\frac{1}{m!\mb{P}_{\mrm{A}}(\mc{N})}$.
Hence, we propose the following algorithm for estimating $|\mc{L}(\bar{d})|$.
\vspace{.1in}
\hrule
\vspace{.07in}
\noindent Algorithm: {\bf CountGraphs}
\vspace{.07in}
\hrule
\vspace{.1in}
\noindent{\tt INPUT:} A graphical degree sequence $\bar{d}$, positive numbers $\ep,\delta$, and an integer $k=k(\ep,\delta)$.\\
\noindent{\tt OUTPUT:} An $(\ep,\delta)$-estimate $X$ for the number of graphs with degree sequence $\bar{d}$.
\vspace{.1in}

\begin{itemize}
\item[(1)] Run Procedure $\mrm{A}$ $k=k(\ep,\delta)$ times, and denote the corresponding values for the random variable $N$ by $N_1,\ldots,N_k$.
\item[(2)] Output $X=\frac{N_1+\cdots+N_k}{k}$ as an estimate for $|\mc{L}(\bar{d})|$.
\end{itemize}
\vspace{.05in}
\hrule \vspace{.1in}
We will show in Section \ref{subsec:varN=o(1)}, that the variance of the random variable $N$ is small enough and therefore, an integer $k=k(\ep,\delta)=O(\ep^{-2}\log(1/\delta))$ exists such that the algorithm CountGraphs produces an $(\ep,\delta)$-estimate for $|\mc{L}(\bar{d})|$.

\subsubsection{Approximating $\mb{P}_{\mrm{A}}(G)$  with SIS.}
Similar to the above discussion, we will use SIS to to find a very close approximation for $\mb{P}_{\mrm{A}}(G)$ for each graph $G$.
Recall that for any graph $G$, each ordering $\mc{N}$ of the edges of $G$ is generated with probability $\mb{P}_{\mrm{A}}(\mc{N})$ by Procedure $\mrm{A}$. Now let $\mrm{S}(G)$ be the set of all $m!$ orderings of $G$. Therefore, the probability $\mb{P}_{\mrm{A}}(G)$ is given by
\begin{equation}
\label{eq:sumP(piG)}
\mb{P}_{\mrm{A}}(G)=\sum_{\mc{N}\in \mrm{S}(G)}\mb{P}_{\mrm{A}}({\mc{N}}).
\end{equation}
Let $\mb{H}$ be the uniform distribution on the set $\mrm{S}(G)$. Then equation \eqref{eq:sumP(piG)} is equivalent to $\mb{P}_{\mrm{A}}(G)=m!~\mb{E}_{\mb{H}}(\mb{P}_{\mrm{A}}({\mc{N}}))$.

Therefore, we use $\mb{H}$ as trial distribution and draw $\ell$ iid samples $\mc{N}_1,\ldots,\mc{N}_\ell$ from $\mb{H}$. Then for each sample $\mc{N}_i$ we calculate $\mb{P}_{\mrm{A}}(\mc{N}_i)$ and report
$$\widehat{\mb{P}_{\mrm{A}}(G)}=\frac{m!}{\ell}\sum_{i=1}^\ell \mb{P}_{\mrm{A}}(\mc{N}_i)$$
as an estimate for $\mb{P}_{\mrm{A}}(G)$. This is given by Procedure $\mrm{B}$.
%----------------------------------------------------------------------------------
\vspace{.1in} \hrule \vspace{.07in}
\noindent Procedure {\bf $\mrm{B}$}
\vspace{.07in} \hrule \vspace{.1in}
\noindent{\tt INPUT:} A graph $G$ with degree sequence $\bar{d}$, and an integer $\ell=\ell(\epsilon,\delta)$.\\
\noindent {\tt OUTPUT:} A real number $P_G$ that is an $(\epsilon,\delta)$-estimate for $\mb{P}_{\mrm{A}}(G)$.
\vspace{.1in}
\begin{itemize}
\item[(1)] Let $E$ be a set of edges, $\hat{d}=(\hat{d}_1,\ldots,\hat{d}_n)$ be an $n$-tuple of
integers, and $P$ be a real number. Initialize them by
$E=$ empty set, $\hat{d}=\bar{d},$ and $P=1$.
\item[(2)] Choose an edge $e=\{v_i,v_j\}$ of $G$ among all those edges that are not in $E$, uniformly at random.
Update $P$ by
$$P=\frac{\hat{d}_i\hat{d}_j(1-\f{d_id_j}{4m})P}{\sum_{\stackrel{(v_r,v_s)\notin E}{v_r\neq v_s}}\hat{d}_r\hat{d}_s(1-\f{d_rd_s}{4m})}~.$$
Add $\{v_i,v_j\}$ to $E$ and reduce each of $\hat{d}_i,~\hat{d}_j$ by $1$.
\item[(3)] Repeat step (2) until $|E|=m$.
\item[(4)] Repeat steps (1) to (3) exactly $\ell=\ell(\ep,\delta)$ times and
let $P_1,\ldots,P_\ell$ be the corresponding values for $P$. Output
$P_G=m!\frac{P_1+\cdots+P_\ell}{\ell}$ as an estimate for $m!~\mb{E}_{\mb{H}}(\mb{P}_{\mrm{A}}(\pi_G))=\mb{P}_{\mrm{A}}(G)$.
\end{itemize}
\vspace{.05in}
\hrule \vspace{.1in}
%----------------------------------------------------------------------------------
Note that the variable $P$ at the end of step (3) is exactly $\mb{P}_{\mrm{A}}(\mc{N})$ for an element  $\mc{N}\in\mrm{S}(G)$ that is sampled from distribution $\mb{H}$.  Therefore, it is easy to see that
$\mb{E}_{\mrm{B}}(P)=\mb{E}_{\mrm{H}}(\mb{P}_{\mrm{A}}(\mc{N}))=\mb{P}_{\mrm{A}}(G)/m!$ which makes $P_G$ an
unbiased estimate for $\mb{P}_{\mrm{A}}(G)$. In Section \ref{subsec:varP=o(1)}, by controlling the variance of the random variable $P$, we will show the existence of an $\ell=\ell(\ep,\delta)=O(\ep^{-2}\log(1/\delta))$ such that the value of $P_G$ is an $(\ep,\delta)$-estimate for $\mb{P}_{\mrm{A}}(G)$ .

\subsubsection{FPRAS for Random Generation.} Now that we can find $(\ep,\delta)$-estimates for both $|\mc{L}(\bar{d})|$ and  $\mb{P}_{\mrm{A}}(G)$ then an FPRAS for random generation is within reach. Algorithm GenerateGraph, given below provides such an FPRAS.
%----------------------------------------------------------------------------------
\vspace{.1in} \hrule \vspace{.07in}
\noindent Algorithm: {\bf GenerateGraph}
\vspace{.07in} \hrule \vspace{.1in}
\noindent {\tt INPUT:} A graphical degree sequence $\bar{d}$ and a positive numbers $\ep$.\\
\noindent {\tt OUTPUT:} A graph $G$ with degree sequence $\bar{d}$.
\vspace{.1in}
\begin{itemize}
\item[(1)] Let $\ep'=\min(0.25,1-\frac{1}{\sqrt{1+\f{\ep}{2}}},\frac{1}{\sqrt{1-\f{\ep}{2}}}-1)$ and $\delta<0.25$.

\item[(2)] Run Algorithm CountGraph, to obtain $X$ as an $(\ep',\delta)$-estimate for $|\mc{L}(\bar{d})|$.

\item[(3)] Repeat Procedure $\mrm{A}$ to obtain one successful outcome $G$.

\item[(4)] Run Procedure $\mrm{B}$ to obtain an $(\ep',\delta)$-estimate, $P_G$, for $\mb{P}_{\mrm{A}}(G)$.

\item[(5)] Report $G$ with probability $\min(\frac{1}{\ct XP_G},1)$ and end. Otherwise go to step (3).
\end{itemize}
\vspace{.05in}
\hrule
\vspace{.1in}
%----------------------------------------------------------------------------------

We will show in Section \ref{sec:analysis} that a universal constant $\ct$ exists (independent of all parameters $m, \bar{d}, \ep$) where the inequality $\ct XP_G\geq 1$ holds whenever $X\geq (1-\ep')|\mc{L}(\bar{d})|$ and $P_G\geq(1-\ep')\mb{P}_{\mrm{A}}(G)$.
Also note that we always assume $0<\ep<1$. Therefore, $\ep'$ is well defined.

%-------------------------------------------------------------------  CORRECTNESS OF ALGORITHM C
\section{Analysis}\label{sec:analysis}
Let us fix a simple graph $G$ with degree sequence $\bar{d}$.  Recall the weighted configuration model from Section \ref{sec:idea} which is equivalent to Procedure $\mrm{A}$.
Denote the set of all perfect matchings on the mini-vertices of $W$ that lead to $G$ by $\mrm{R}(G)$. Any two
elements of $\mrm{R}(G)$ can be obtained from one another by permuting the labels of the mini-vertices in any $W_i$.
Due to this symmetry, all matchings in $\mrm{R}(G)$ are generated with equal probability using Procedure $\mrm{A}$. In other words for
a fixed element $\mc{M}$ in $\mrm{R}(G)$ we have $\mb{P}_{\mrm{A}}(G)=\left(\prod_{i=1}^n
d_i!\right)\mb{P}_{\mrm{A}}(\mc{M})$.

Now we will find $\mb{P}_{\mrm{A}}(\mc{M})$. First note that there are $m!$
different orders for picking the edges of $\mc{M}$ sequentially. Moreover, different orderings can have different
probabilities. Denote the set of these orderings by $\mrm{S}(\mc{M})$. Thus
\[\mb{P}_{\mrm{A}}(G)=\left(\prod_{i=1}^n d_i!\right)\sum_{\mc{N}\in \mrm{S}(\mc{M})}\mb{P}_{\mrm{A}}(\mc{N}).
\]
For
any ordering $\mc{N}=\{e_1,\ldots,e_m\}$ in the set $\mrm{S}(\mc{M})$ and
each $r$ with $0\leq r\leq m-1$ denote the probability of picking edge $e_{r+1}$ at step $r+1$ of Procedure $\mrm{A}$
by $\mb{P}\left(e_{r+1}|e_1,\ldots,e_r\right)$. Hence
$\mb{P}_{\mrm{A}}(\mc{N})=\prod_{r=0}^{m-1}\mb{P}\left(e_{r+1}|e_1,\ldots,e_r\right)$ and each
term $\mb{P}\left(e_{r+1}|e_1,\ldots,e_r\right)$ is given by
\begin{equation}
\label{eq:P(e_r+1|e1...)}
\mb{P}\Big(e_{r+1}=\{i,j\}|e_1,\ldots,e_r\Big)=\f{(1-d_id_j/4m)}{\sum_{\{u,v\}\in
E_r}d_u^{(r)}d_v^{(r)}(1-d_ud_v/4m)}
\end{equation}
where $d_i^{(r)}$ denotes the residual degree of vertex $i$ at step $r+1$ and the set $E_r$ consists of all possible edges after picking
$e_1,\ldots,e_r$. Note that $d_i^{(r)}$ is also equal to the number of unmatched mini-vertices in $W_i$ at step $r+1$.
For the analysis we use the notations $\{i,j\}$ and $\{v_i,v_j\}$ interchangeably.

{Denote the number of unsuitable pairs after choosing the edges in $\mc{N}_r=\{e_1,\ldots,e_r\}$ by $\Delta_r(\mc{N})$.}
Thus, the denominator of the right hand side of \eqref{eq:P(e_r+1|e1...)} can be written as ${2m-2r\choose2}-\Psi_r(\mc{N})$
where $\Psi_r(\mc{N})=\Delta_r(\mc{N})+\sum_{\{u,v\}\in
E_r}d_u^{(r)}d_v^{(r)}d_ud_v/4m$.
This is because $\sum_{\{u,v\}\in
E_r}d_u^{(r)}d_v^{(r)}$ is the number of the suitable pairs at step $r+1$,
and is equal to $ {2m-2r\choose2}-\Delta_r(\mc{N})$. The quantity
$\Psi_r(\mc{N})$ can be also viewed as sum of the weights of
the unsuitable pairs. Now using $1-x=e^{-x+O(x^2)}$ for $0\leq x\leq 1$, when $d_{\max}=O(m^{1/4-\tau})$ the expression for $\mb{P}_{\mrm{A}}(G)$ is
\begin{eqnarray*}
\mb{P}_{\mrm{A}}(G)&=&\left(\prod_{i=1}^n d_i!\right)\left[\prod_{i\sim_G
j}(1-\f{d_id_j}{4m})\right]\sum_{\mc{N}\in
\mrm{S}(\mc{M})}\prod_{r=0}^{m-1}\f{1}{{2m-2r\choose2}-\Psi_r(\mc{N})}\\
&=&\left(\prod_{i=1}^n d_i!\right)\,e^{-\gamma_G+o(1)}\sum_{\mc{N}\in
\mrm{S}(\mc{M})}\prod_{r=0}^{m-1}\f{1}{{2m-2r\choose2}-\Psi_r(\mc{N})}
\end{eqnarray*}
where $\gamma_G$ was defined in Section \ref{sec:idea} to be $\gamma_G=\sum_{i \sim_G j}d_id_j/4m$.
The next step is to show that  $\Psi_r(\mc{N})$ is sharply concentrated around a number $\psi_r(G)$, independent of the
ordering $\mc{N}$. More specifically for
$$\psi_r(G)=(2m-2r)^2\left(\f{\la(\bar{d})}{2m}+\f{r\sum_{i\sim_Gj}(d_i-1)(d_j-1)}{4m^3}
+\f{(\sum_{i=1}^nd_i^2)^2}{32m^3}+o(1)\right)$$
the following is true
\begin{equation}
\sum_{\mc{N}\in
\mrm{S}(\mc{M})}\prod_{r=0}^{m-1}\f{1}{{2m-2r\choose2}-\Psi_r(\mc{N})}=\left[1+o(1)\right]m!\prod_{r=0}^{m-1}\f{1}{{2m-2r\choose2}-\psi_r(G)}.\label{eq:kimvuconc}
\end{equation}
The proof of this concentration result uses Kim and Vu's polynomial method
\cite{Kim-Vu2000} and is quite technical. It generalizes Kim and Vu's
\cite{Kim-Vu} calculations for the regular graphs to the general degree sequences. Section \ref{sec:Kim-Vu-Calc} is dedicated to this cumbersome analysis. But for the case of regular graphs, in Section \ref{subsec:pf-reg}, we will use a different technique based on Azuma's inequality to show concentration in a larger region.

The next step is to show that when $d_{\max}=O(m^{1/4-\tau})$,
\begin{equation}
\prod_{r=0}^{m-1}\f{1}{{2m-2r\choose2}-\psi_r(G)}=\prod_{r=0}^{m-1}\f{1}{{2m-2r\choose2}}e^{\la(\bar{d})+\la^2(\bar{d})+\gamma_G+o(1)}.
\label{eq:alg}
\end{equation}
The proof of equation (\ref{eq:alg}) is algebraic and is given in
Section \ref{subsec:alg-proof}.

The above analysis can now be summarized in the following lemma.
\begin{lemma}\label{lem:main-general-conc}
For $d_{\max}=O(m^{1/4-\tau})$, Procedure $\mrm{A}$ generates all
graphs with degree sequence $\bar{d}$ with asymptotically equal
probability. More specifically
\begin{equation*}
\sum_{\mc{N}\in
\mrm{S}(\mc{M})}\mb{P}_{\mrm{A}}(\mc{N})=\f{m!}{\prod_{r=0}^m{2m-2r\choose
2}}e^{\la(\bar{d})+\la^2(\bar{d})+o(1)}.%\label{eq:main}
\end{equation*}
\end{lemma}
Now we can prove the first theorem.
\begin{proof}[of Theorem \ref{thm:main-nobias}]
Lemma \ref{lem:main-general-conc} shows that
$\mb{P}_{\mrm{A}}(G)$ is asymptotically independent of $G$.  Therefore, we only need to show
Procedure $\mrm{A}$ always succeeds with probability $1-o(1)$. We
will show this in Section \ref{sec:prob-fail}
%of \cite{BKS07extended}
by proving the following lemma.
\begin{lemma}\label{lem:prob-fail}
For $d_{\max}=O(m^{1/4-\tau})$, the probability of failure of Procedure $\mrm{A}$ is $o(1)$.
\end{lemma}
Therefore, all graphs $G$ are generated with asymptotically uniform
probability. Note that this fact, combined with equation \eqref{eq:kimvuconc} will also give an independent proof of
McKay's formula \cite{Mckay} for the number of graphs.

Finally we are left with the analysis of the running time which is summarized in the following lemma. The proof of this lemma is given in Section \ref{sec:procB-run-time}.
\begin{lemma}\label{lem:rtime}
Procedure $\mrm{A}$ can be implemented so that the expected running time is O($m\,d_{\max}$) for $d_{\max}=O(m^{1/4-\tau})$.
\end{lemma}
This completes the proof of Theorem \ref{thm:main-nobias}.
\enp\end{proof}

\subsubsection{Proof of Theorem \ref{thm:fpras}.}

First we will prove that Algorithm CountGraphs is an FPRAS for the counting problem. This is shown by the following lemma.
\begin{lemma}\label{lem:PUestimate}
For any $\ep,\delta>0$ there exist
$k=k(\ep,\delta)=O(\epsilon^{-2}\log(1/\delta))$ such that the output $X$ of Algorithm CountGraphs is an
$(\ep,\delta)$-estimate for $|\mc{L}(\bar{d})|$.
\end{lemma}
\begin{proof}
Since $\mb{E}_{\mrm{A}}(N)=\mc{L}(\bar{d})$,
\begin{multline}
\mb{P}\Big[(1-\epsilon)|\mc{L}(\bar{d})|<X<(1+\epsilon)|\mc{L}(\bar{d})|\Big]\\
=\mb{P}\left(-\frac{\epsilon\mb{E}_{\mrm{A}}(N)}{\sqrt{\frac{\textrm{Var}_{\mrm{A}}(N)}{k}}}<
\f{X-\mb{E}_{\mrm{A}}(X)}{\sqrt{\frac{\textrm{Var}_{\mrm{A}}(N)}{k}}}<\frac{\epsilon\mb{E}_{\mrm{A}}(N)}{\sqrt{\frac{\textrm{Var}_{\mrm{A}}(N)}{k}}}\right)
\end{multline}
On the other hand, as a consequence of the Central Limit Theorem, when $k$ goes to infinity, the quantity
$\f{X-\mb{E}_{\mrm{A}}(X)}{\sqrt{\textrm{Var}_{\mrm{A}}(N)/k}}$ converges to a random variable $Z$
which has a normal distribution with mean zero and variance $1$.
Therefore similar to the discussion given in \cite{blanchet}, the inequality
$\frac{\epsilon\mb{E}_{\mrm{A}}(N)}{\sqrt{\textrm{Var}_{\mrm{A}}(N)/k}}>z_{\delta}$
guarantees that $X$ is an $(\ep,\delta)$-estimate for $|\mc{L}(\bar{d})|$
where $\mb{P}(|Z|>z_{\delta})=\delta$. This condition is equivalent
to the following lower bound for the number of repetitions of Procedure $\mrm{A}$
$$k>z_\delta^2\epsilon^{-2}\frac{\textrm{Var}_{\mrm{A}}(N)}{\mb{E}_{\mrm{A}}(N)^2}.$$
Moreover, the tail of the normal distribution, $\mb{P}(|Z|>x)$, for very large values of $x$ can be approximated
by the quantity $ax^{-1}e^{-x^2/2}(2\pi)^{-1}$ where $a>0$ is a constant. This means that the quantity $z_\delta^2$ is of $O(\log(1/\delta))$.
Therefore, if we show that the variance ratio $\textrm{Var}_{\mrm{A}}(N)/\mb{E}_{\mrm{A}}(N)^2$ is bounded from above by a constant, then with $k = O(\log(1/\delta)\epsilon^{-2})$ repetitions, we can  obtain an $(\ep,\delta)$-estimate. In fact we will prove the stronger statement
\begin{equation}\frac{\textrm{Var}_{\mrm{A}}(N)}{\mb{E}_{\mrm{A}}(N)^2}=o(1)\label{eq:varN=o(1)}
\end{equation}
in Section \ref{subsec:varN=o(1)}.
%of \cite{BKS07extended}.
This finishes the proof of Lemma \ref{lem:PUestimate}.
\enp\end{proof}
Note that By Theorem \ref{thm:main-nobias}, Procedure $\mrm{A}$
uses $O(m\,d_{\max})$ operations. Therefore the running time of Algorithm
CountGraphs is $k(\ep,\delta)$ times $O(m\,d_{\max})$ which is
$O(m\,d_{\max}\ep^{-2}\log(1/\delta))$. This shows that the algorithm  CountGraphs is an FPRAS for estimating $|\mc{L}(\bar{d}|$.

Now we will prove that Algorithm GenerateGraph is an FPRAS for the random generation problem as well.
First notice that if the ratio $\textrm{Var}_{\mrm{B}}(P)/\mb{E}_{\mrm{B}}(P)^2$ is bounded from above by a constant, then similar calculations as in the proof of Lemma \ref{lem:PUestimate} for the tail of the normal distribution can be used to find $\ell=\ell(\ep,\delta)=O(\ep^{-2}\log(1/\delta))$ such that the output of Procedure $\mrm{B}$, $P_G$, is an $(\ep,\delta)$-estimate for $\mb{P}_{\mrm{A}}(G)$. In fact we will show the stronger result
\begin{equation}
\frac{\textrm{Var}_{\mrm{B}}(P)}{\mb{E}_{\mrm{B}}(P)^2}= o(1)\label{eq:varP=o(1)}
\end{equation}
in Section \ref{subsec:varP=o(1)}.
Therefore, equation \eqref{eq:varP=o(1)} gives the following lemma.
\begin{lemma}\label{lem:ProcB}
For any $\epsilon,\delta>0$ and a graph $G$ with degree sequence $\bar{d}$, there exist $\ell=\ell(\ep,\delta)=O(\ep^{-2}\log(1/\delta))$ for
Procedure $\mrm{B}$ such that its output, $P_G$, is an $(\ep,\delta)$-estimate for $\mb{P}_{\mrm{A}}(G)$.
\end{lemma}

The next step in analyzing Algorithm GenerateGraph is to prove the existence of constant $\ct$ that is used in Step $(5)$.
\begin{lemma}\label{lem:contant-c}
There exists a constant $\ct$ such that for all parameters $m,\bar{d},\ep$ and all graphs $G$ with degree sequence $\bar{d}$, the inequality $\ct XP_G\geq 1$  holds whenever $X\geq (1-\ep')|\mc{L}(\bar{d})|$ and $P_G\geq (1-\ep')\mb{P}_{\mrm{A}}(G)$.
\end{lemma}
\begin{proof}
By Theorem \ref{thm:main-nobias}, $\big[1-o(1)\big]|\mc{L}(\bar{d})|^{-1}\leq\mb{P}_{\mrm{A}}(G)\leq \big[1+o(1)\big]|\mc{L}(\bar{d})|^{-1}$.
Let $M$ be large enough such that for all $m>M$ the $o(1)$ terms are less than $1/2$.  Now define
$$\dt=\min\left(\frac{1}{2}~,\min_{m\leq M}\min_{G\in \mc{L}(\bar{d})}(\mb{P}_{\mrm{A}}(G)|\mc{L}(\bar{d})|)\right)$$
and
$$\et=\max\left(\frac{3}{2}~,\max_{m\leq M}\max_{G\in \mc{L}(\bar{d})}(\mb{P}_{\mrm{A}}(G)|\mc{L}(\bar{d})|)\right).$$
Therefore, $\dt$ and $\et$ are positive and finite constants that are independent of all of the parameters $m,\bar{d},\ep$ and
$\dt\leq\mb{P}_{\mrm{A}}(G)|\mc{L}(\bar{d})|\leq \et$.
Now when $X\geq (1-\ep')|\mc{L}(\bar{d})|$ and $P_G\geq (1-\ep')\mb{P}_{\mrm{A}}(G)$,
\begin{equation*}
\frac{\dt}{4}\leq \dt(1-\ep')^2\leq P_GX.
\end{equation*}
This is because $\ep'\leq 0.25$. Therefore $\ct = {4}/{\dt}$ suffices.
\enp
\end{proof}
Now we need to analyze the output distribution and the running time of Algorithm GenerateGraph.
Consider one iteration of Algorithm GenerateGraph from step (1) to step (5).  Let $Ev_1$ be the event that at least one of the fractions $\frac{X}{|\mc{L}(\bar{d})|}$ or $\frac{P_G}{\mb{P}_{\mrm{A}}(G)}$ is not in the interval $[1-\ep',1+\ep']$. Let $Ev_2$ be the event that a graph is reported in step (5).  This means $Ev_2^c$ is when "Otherwise go to step (3)" is called.
Therefore, $\mb{P}(Ev_1)\leq 2\delta<0.5$ and $\mb{P}(Ev_1)+\mb{P}(Ev_2|Ev_1^c)\mb{P}(Ev_1^c)+\mb{P}(Ev_2^c|Ev_1^c)\mb{P}(Ev_1^c)=1$.

For each graph $G\in \mc{L}(\bar{d})$ let $Ev_2(G)$ be the event that $G$ is reported in step (5).  Each graph $G$ is reported with probability $\mb{P}(Ev_2(G)|Ev_1^c)=\mb{P}_{\mrm{A}}(G)/(\ct XP_G)$ that satisfies
\begin{equation}
\label{eq:1pmeps.1/cL}
\frac{1-\ep/2}{\ct|\mc{L}(\bar{d})|}\leq\frac{1}{\ct(1+\ep')^2|\mc{L}(\bar{d})|}\leq \mb{P}(Ev_2(G)|Ev_1^c) \leq \frac{1}{\ct(1-\ep')^2|\mc{L}(\bar{d})|}\leq \frac{1+\ep/2}{\ct|\mc{L}(\bar{d})|}.
\end{equation}
Note that the events $Ev_2(G)|Ev_1$ are not important and have low probability. Now we obtain
$$\mb{P}(Ev_2) \geq \mb{P}(Ev_1^c)\sum_{G\in \mc{L}(\bar{d})}\mb{P}(Ev_2(G)|Ev_1^c)\geq \frac{0.5(1-\ep/2)}{\ct}.$$
Therefore, the expected number of times that ``Otherwise go to Step $(3)$'' is called is $\mb{P}(Ev_2)^{-1}\leq 4\ct$.  This means that the expected running time of Algorithm GenerateGraph is at most $4\ct$ times the expected running time of a successful run of Procedure $\mrm{A}$ plus $4\ct$ times the expected running time of Procedure $\mrm{B}$ plus the expected running time of Algorithm CountGraphs. The total number of operations can be written as
$$4\ct O(md_{\max})+4\ct O(md_{\max}\ep'^{-2}\log(1/\delta))+ O(md_{\max}\ep'^{-2}\log(1/\delta))$$
which is $O(md_{\max}\ep^{-2})$, since $\ep'\geq \min(\ep/4,0.25)$ gives $\ep'^{-2}=O(\ep^{-2})$.

Notice that the probability that Algorithm GenerateGraph eventually reports a graph, in an iteration that $Ev_1$ did not occur, is
at least $1-\mb{P}(Ev_1)> 0.5$.  Moreover, the probability that the reported graph is a fixed graph $G\in \mc{L}(\bar{d})$ satisfies
$$\sum_{i=0}^{\infty}\mb{P}(Ev_2^c)^i\mb{P}(Ev_2(G)|Ev_1^c)\mb{P}(Ev_1^c)=\frac{\mb{P}(Ev_2(G)|Ev_1^c)\mb{P}(Ev_1^c)}{\mb{P}(Ev_2)}\in [\f{1-\ep}{c'|\mc{L}(\bar{d})|},\f{1+\ep}{c'|\mc{L}(\bar{d})|}]$$
where $c'=\frac{\mb{P}(Ev_1^c)}{\ct\mb{P}(Ev_2)}$. This finishes the proof of Theorem \ref{thm:fpras}.
\enp

\subsection{Concentration inequality for regular graphs}\label{subsec:pf-reg}

The aim of this section is to prove Theorem \ref{thm:main-regular}. Recall that $\mc{L}(n,d)$ denotes the set of all simple $d$-regular
graphs with $m=nd/2$ edges.
Let $\mb{P}_{\mrm{U}}$ be the uniform probability on $\mc{L}(n,d)$.
Similar to the analysis of Procedure $\mrm{A}$ for general degree sequences, let $G$ be a fixed
graph in $\mc{L}(n,d)$ and $\mc{M}$ be a fixed matching on $W$ with $G_{\mc{M}}=G$.
The main goal is to show that for
$d=o(n^{1/2-\tau})$ the probability of generating $G$ with Procedure $\mrm{A}$ is at least
$1-o(1)$ times $\mb{P}_{\mrm{U}}(G)$; i.e.
\begin{equation}
\label{eq:PA(reg)>=uniform}
\mb{P}_{\mrm{A}}(G)\geq \big(1-o(1)\big)\mb{P}_{\mrm{U}}(G).
\end{equation}
For the moment, assume \eqref{eq:PA(reg)>=uniform} is true. We will show that Theorem \ref{thm:main-regular} follows. Later we will show why \eqref{eq:PA(reg)>=uniform} holds.
\begin{proof}[of Theorem \ref{thm:main-regular}]
First, we will show that the total variation distance between the probability measures $\mb{P}_{\mrm{A}}$ and $\mb{P}_{\mrm{U}}$,
$\dv(\mb{P}_{\mrm{A}}, \mb{P}_{\mrm{U}}) \equiv \sup_{\mc{S}\subset \mc{L}(n,d)}\,|\mb{P}_{\mrm{A}}(\mc{S})- \mb{P}_{\mrm{U}}(\mc{S})|$ is $o(1)$.
We will use the following upper bound on the total variation distance
\[\dv(\mb{P}_{\mrm{A}}, \mb{P}_{\mrm{U}}) \leq \sum_{G\in \mc{L}(n,d)}\,|\mb{P}_{\mrm{A}}(G)- \mb{P}_{\mrm{U}}(G)|.\]
Therefore, we have the upper-bound
\begin{eqnarray}
 \sum_{G\in \mc{L}(n,d)}\,|\mb{P}_{\mrm{A}}(G)- \mb{P}_{\mrm{U}}(G)|&=&\sum_{\stackrel{G\in \mc{L}(n,d)}{\mb{P}_{\mrm{A}}\geq \mb{P}_{\mrm{U}}}}\Big(\mb{P}_{\mrm{A}}(G)- \mb{P}_{\mrm{U}}(G)\Big)
+ \sum_{\stackrel{G\in \mc{L}(n,d)}{\mb{P}_{\mrm{A}}<\mb{P}_{\mrm{U}}}}|\mb{P}_{\mrm{A}}(G)- \mb{P}_{\mrm{U}}(G)|\no\\
&=&\sum_{G\in \mc{L}(n,d)}\Big(\mb{P}_{\mrm{A}}(G)- \mb{P}_{\mrm{U}}(G)\Big)
+ 2\sum_{\stackrel{G\in \mc{L}(n,d)}{\mb{P}_{\mrm{A}}<\mb{P}_{\mrm{U}}}}|\mb{P}_{\mrm{A}}(G)- \mb{P}_{\mrm{U}}(G)|\no\\
&\stackrel{(a)}{\leq}&2\sum_{\stackrel{G\in \mc{L}(n,d)}{\mb{P}_{\mrm{A}}<\mb{P}_{\mrm{U}}}}|\mb{P}_{\mrm{A}}(G)- \mb{P}_{\mrm{U}}(G)|\no\\
&\stackrel{(b)}{\leq}&2\,o(1)\sum_{\stackrel{G\in \mc{L}(n,d)}{\mb{P}_{\mrm{A}}<\mb{P}_{\mrm{U}}}}\mb{P}_{\mrm{U}}(G)~\leq~ o(1).\no
\end{eqnarray}
Here $(a)$ uses $\sum_{G\in \mc{L}(n,d)}\mb{P}_{\mrm{A}}(G)\leq 1$ and $\sum_{G\in \mc{L}(n,d)}\mb{P}_{\mrm{U}}(G)=1$.
To see why $(b)$ holds, note that $\mb{P}_{\mrm{U}}(G)-\mb{P}_{\mrm{A}}(G)\leq o(1)\mb{P}_{\mrm{U}}(G)$ which is equivalent to inequality \eqref{eq:PA(reg)>=uniform}.

Now, $\dv(\mb{P}_{\mrm{A}}, \mb{P}_{\mrm{U}})=o(1)$ implies that $\mb{P}_{\mrm{A}}(G)\leq \big(1+o(1)\big)\mb{P}_{\mrm{U}}(G)$ except for graphs $G$ in a subset of $\mc{L}(n,d)$ with size $o(|\mc{L}(n,d)|)$. This finishes the proof of Theorem \ref{thm:main-regular}.
\enp
\end{proof}

\subsubsection{Proof of inequality \eqref{eq:PA(reg)>=uniform}.}
In order to prove inequality \eqref{eq:PA(reg)>=uniform} we prove the following equivalent inequality
\begin{equation}
\label{eq:PA(reg)>=uniform2}
(d!)^n\sum_{\mc{N}\in
\mrm{S}(\mc{M})}\mb{P}(\mc{N})\geq
\frac{
1-o(1)}{|\mc{L}(n,d)|}.
\end{equation}
Our proof of inequality \eqref{eq:PA(reg)>=uniform2} builds upon the steps
in Kim and Vu \cite{KimVuSandwitch}. First define $\mu_r=\mu_r^{(1)}+\mu_r^{(2)}$ where
\begin{eqnarray*}
\mu_r^{(1)}&=&\frac{(2m-2r)^2(d-1)}{4m}\\
\mu_r^{(2)}&=&\frac{(2m-2r)^2(d-1)^2r}{4m^2}.
\end{eqnarray*}
Let $m_1=\f{m}{d^2\om}$ where $\om$ goes to
infinity very slowly; e.g. $\om=O(\log^\delta n)$ for some small $\delta>0$.
The following summarizes the
analysis of Kim and Vu \cite{KimVuSandwitch} for
$d=O(n^{1/3-\tau})$
\begin{eqnarray}
|\mc{L}(n,d)|(d!)^n\sum_{\mc{N}\in \mrm{S}(\mc{M})}\mb{P}(\mc{N})
&\stackrel{(c)}{=}&\f{1-o(1)}{m!}\sum_{\mc{N}\in
\mrm{S}(\mc{M})}\prod_{r=0}^{m-1}\f{{2m-2r\choose2}-\mu_r}{{2m-2r\choose2}-\Delta_r(\mc{N})}\no\\
&\stackrel{(d)}{\geq}&\f{1-o(1)}{m!}\sum_{\mc{N}\in
\mrm{S}(\mc{M})}\prod_{r=0}^{m_1}\left(1+\f{\Delta_r(\mc{N})-\mu_r}{{2m-2r\choose2}-\Delta_r(\mc{N})}\right)\no\\
&\stackrel{(e)}{\geq}&\Big(1-o(1)\Big)\prod_{r=0}^{m_1}\left(1-3\f{T_r^{(1)}+T_r^{(2)}}{(2m-2r)^2}\right)\no\\
&\stackrel{(f)}{\geq}&\Big(1-o(1)\Big)\exp\left(-3e\sum_{r=0}^{m_1}\f{T_r^{(1)}+T_r^{(2)}}{(2m-2r)^2}\right).
\label{eq:main-reg}
\end{eqnarray}
Here we explain these steps in more detail. Our main focus will be on step $(e)$ which is the main step. For the rest, we provide a brief description and a reference to \cite{KimVuSandwitch}.
Step $(c)$ follows from equation
$(3.5)$ of \cite{KimVuSandwitch} and writing McKay-Wormald's estimate
\cite{MckayWormald1991} for $|\mc{L}(n,d)|$ as a multiple of the product $\prod_{r=0}^{m-1}\left[{2m-2r\choose2}-\mu_r\right]$.
Similarly, step $(d)$ follows from the algebraic calculations in page 455 of \cite{KimVuSandwitch}.

The important step $(e)$ follows from a sharp concentration. For simplicity
write $\Delta_r$ instead of $\Delta_r(\mc{N})$ and break
$\Delta_r$ into two terms $\Delta_r^{(1)}+\Delta_r^{(2)}$. Here $\Delta_r^{(1)}$ and
$\Delta_r^{(2)}$ denote the number of unsuitable pairs in step $r$
corresponding to the self edge loops and to the double edges respectively. For
$p_r=r/m$, $q_r=1-p_r$ Kim and Vu \cite{KimVuSandwitch} used their
polynomial concentration inequality \cite{Kim-Vu2000} to derive two bounds $T_r^{(1)},~
T_r^{(2)}$ and to show that with with very high probability
$|\Delta_r^{(1)}-\mu_r^{(1)}|<T_r^{(1)}$ and $|\Delta_r^{(2)}-\mu_r^{(2)}|<T_r^{(2)}$. More
precisely for some constants $c_1,c_2$ the bounds are
\[T_r^{(1)}=c_1\log^2 n\sqrt{nd^2q_r^2(2dq_r+1)}~~~~,~~~~T_r^{(2)}= c_2\log^3
n\sqrt{nd^3q_r^2(d^2q_r+1)}.\]
Now it is easy to see that for each $i\in\{1,2\}$ the bound $T_r^{(i)}$ and the quantity $\mu_r^{(i)}$ are $o\left((2m-2r)^2\right)$. This validates the step
$(e)$.

Finally, the step $(f)$ is straightforward using $1-x\geq e^{-ex}$ for $0\leq x\leq 1$.

The rest of the proof focuses on showing that the right hand side of inequality \eqref{eq:main-reg}
is at least $1-o(1)$.  Kim and Vu show that for $d=O(n^{1/3-\tau})$ the exponent in equation
(\ref{eq:main-reg}) is $o(1)$. Using similar calculations as
equation (3.13) in \cite{KimVuSandwitch} it can be shown that for
$d=O(n^{1/2-\tau})$ and $m_2=(m\log^3n)/d$
\[\sum_{r=0}^{m_1}\f{T_r^{(1)}}{(2m-2r)^2}=o(1)~~~~,~~~~\sum_{r=m_2}^{m_1}\f{T_r^{(2)}}{(2m-2r)^2}=o(1).\]
But unfortunately the summation
$\sum_{r=0}^{m_2}\f{T_r^{(2)}}{(2m-2r)^2}$ is $\Om(d^3/n)$. In
fact it turns out that the random variable $\Delta_r^{(2)}$ has large
variance for $d=O(n^{1/2-\tau})$.

Let us explain the main difficulty for moving from
$d=O(n^{1/3-\tau})$ to $d=O(n^{1/2-\tau})$. Note that $\Delta_r^{(2)}$
is defined on a random subgraph $G_{\mc{N}_r}$ of graph $G$
which has exactly $r$ edges. Both \cite{StegerWormald} and
\cite{Kim-Vu,KimVuSandwitch} have approximated the subgraph $G_{\mc{N}_r}$
with $G_{p_r}$ in which each edge of $G$ appears independently with
probability $p_r=r/m$. But when
$d=O(n^{1/2-\tau})$, this approximation causes the variance of
$\Delta_r^{(2)}$ to become exponentially large.

In order to fix the problem, we modify $\Delta_r^{(2)}$ before moving to
$G_{p_r}$. It can be shown via simple algebraic calculations that: $
\Delta_r^{(2)} -\mu_r^{(2)} = X_r - Y_r$ where
\begin{eqnarray*}
X_r&=&\sum_{u\sim_{G_{\mc{N}_r}} v}
[d_u^{(r)}-q_r(d-1)][d_v^{(r)}-q_r(d-1)]\\
Y_r&=&q_r(d-1)\sum_u
\left[(d_u^{(r)}-q_rd)^2-dp_rq_r\right].
\end{eqnarray*}
This modification is
critical since the equality $ \Delta_r^{(2)} -\mu_r^{(2)} = X_r - Y_r$ does
not hold in $G_{p_r}$.

The next task is to find a new bound $\hat{T}_r^{(2)}$ such that
$|X_r-Y_r|<\hat{T}_r^{(2)}$ with very high probability and
$\sum_{r=0}^{m_2}\f{\hat{T}_r^{(2)}}{(2m-2r)^2}=o(1)$. It is easy to see
that in $G_{p_r}$ both $X_r$ and $Y_r$ have zero expected value.

At this time we will move to $G_{p_r}$ and show that $X_r$ and $Y_r$
are sharply concentrated around zero. It is easy to see that with probability at least $1/n$, the subgraph $G_{p_r}$ has exactly $r$
edges. This is in fact Lemma \ref{lem:Gp-numedge} which is proven in Section \ref{sec:Kim-Vu-Calc}.
Therefore, $X_r$ and $Y_r$ will be sharply concentrated around $0$ in $G_{\mc{N}_r}$ as well.
In the following we will show the concentration of $X_r$ in $G_{p_r}$. The concentration of $Y_r$ can be shown similarly.

Consider the edge exposure martingale (page 94 of \cite{AS}) for $G_{p_r}$ that examines the edges of $G$ in the order $e_1,\ldots,e_m$. In particular
for any $0\leq\ell\leq r$ define $Z_{\ell}^r = \mb{E}(X_r~|~e_1,\ldots,e_\ell)$. Therefore, $Z_m^r$ is just the value of $X_r$ and $Z_0^r$ is its expected value $\mb{E}(X_r)$ in $G_{p_r}$.  To simplify the notation, let us drop the index $r$ from $Z_\ell^r,
d_u^{(r)}, p_r$ and $q_r$.

The next step is to bound the martingale difference $|Z_i-Z_{i-1}|$ and use a martingale concentration inequality.
In order to bound the quantity $|Z_i-Z_{i-1}|$, assume that $e_i=\{u,v\}$. The difference between $Z_i$ and $Z_{i-1}$ is in the terms involving $e_i$ in the summation $\sum_{u'\sim_{G_p} v'} [d_{u'}-q(d-1)][d_{v'}-q(d-1)]$. But $e_i$ only participates in $d_u$ and $d_v$. Thus, for any $u'$ where $u'\sim_{G_p} u$, the term $[d_{u'}-q(d-1)][d_{u}-q(d-1)]$ appears in both $Z_i$ and $Z_{i-1}$. The value of $d_{u'}-q(d-1)$ is unchanged by revealing the status of $e_i$, but the value of $d_{u}-q(d-1)$ can fluctuate by at most $1$.  Moreover, if $e_i\in G_p$ then an extra term $[d_u-q(d-1)][d_v-q(d-1)]$ is also added to $Z_i$. This means we have
\begin{multline}
\label{eq:mart-diff}
|Z_i-Z_{i-1}|\leq
\bigg|\big(d_u-(d-1)q\big)\big(d_v-(d-1)q\big)\bigg|\\
+\bigg|\sum_{u'\sim_{G_p} u}\big(d_{u'}-(d-1)q\big)\bigg|+\bigg|\sum_{v'\sim_{G_p}
v}\big(d_{v'}-(d-1)q\big)\bigg|~.
\end{multline}
Bounding the above difference should be done carefully since the
standard worst case bounds are weak for our purpose.

First, we start by a useful observation. For a typical ordering $\mc{N}$ of the edges of $G$,
the residual degrees, $d_u,~d_v,~d_{u'},~d_{v'}$ are roughly $dq\pm\sqrt{dq}$. We
will make this more precise. For any vertex $\bar{u}\in G$ consider the event
$$L_{\bar{u}}=\big\{|d_{\bar{u}}-dq|\leq c\log^{1/2}n(dq)^{1/2}\big\}$$
where $c>0$
is a large constant.
\begin{lemma}\label{lem:d-sqrtd}
For all $0\leq r\leq m_2$ we have $\mb{P}(L_{\bar{u}}^c)=o(\f{1}{m^4})$.
\end{lemma}
\begin{proof} Note that in the $G_p$
model the residual degree of a vertex $\bar{u}$, $d_{\bar{u}}$, is sum of $d$
independent Bernoulli random variables with mean $q$. Two
generalizations of Chernoff inequality (Theorems A.1.11, A.1.13
in page 267 of \cite{AS}) state that for $a>0$ and $X_1,\ldots,X_d$
i.i.d. Bernoulli(q) random variables:
\begin{eqnarray*}
\mb{P}(X_1+\cdots+X_d-qd\geq a)&<&e^{-\f{a^2}{2qd}+\f{a^3}{2(qd)^2}}\\
\mb{P}(X_1+\cdots+X_d-qd< -a)&<&e^{-\f{a^2}{2qd}}
\end{eqnarray*}
Applying these two for $a=\sqrt{12qd\log n}$ proves Lemma \ref{lem:d-sqrtd}.
\enp\end{proof}
To finish bounding the martingale difference we look at the last two terms
in the right hand side of equation (\ref{eq:mart-diff}). For the vertex $u$ consider the
event
$$K_u=\Big\{\big|\sum_{u'\sim_{G_p} u}(d_{u'}-(d-1)q)\big|\leq
c\big[(dq)^{3/2}+qd+dq^{1/2}\big]\log n\Big\}$$
where $c>0$ is
a large constant. We will use the following lemma to show that the complement of $K_u$ has very low probability.
\begin{lemma}\label{lem:d-sqrtd-2}
For all $0\leq r\leq m_2$ the event $K_u^c$ has probability $o(\f{d}{m^4})$.
\end{lemma}
\begin{proof}
For any vertex $u$ let
$N_G(u)\subset V(G)$ denote the neighbors of $u$ in $G$. Consider the subsets
$$A_G(u),~B_G(u),~C_G(u)\subset E(G)$$
where $A_G(u)$ consists of the edges that are adjacent to $u$, $B_G(u)$ has those edges with both
endpoints in $N_G(u)$, and $C_G(u)$ contains those edges with
exactly one endpoint in $N_G(u)$ and one endpoint outside $N_G(u)\cup\{u\}$. For any edge $e$ of $G$ let
$t_e=\mrm{1}_{\{e\notin G_p\}}$. Then we can write
\begin{eqnarray}
&&\sum_{u'\sim_{G_p} u}\big(d_{u'}-(d-1)q\big)\no\\
&=&\sum_{u'\in N_{G_p}(u)}~~\sum_{e\in A_G(u')\setminus A_G(u)}(t_e-q)\no\\
&=&\sum_{u'\in N_G(u)}~~\sum_{e\in A_G(u')\setminus
A_G(u)}(t_e-q)-\sum_{u'\in N_G(u)\setminus N_{G_p}(u)}~~\sum_{e\in A_G(u')\setminus A_G(u)}(t_e-q)\no\\
&=&\underbrace{\sum_{e\in C_G(u)}(t_e-q)}_{(i)}
+2\underbrace{\sum_{e\in B_G(u)}(t_e-q)}_{(ii)}
-\underbrace{\sum_{u'\in N_G(u)\setminus
N_{G_p}(u)}\big(d_u'-1-q(d-1)\big)}_{(iii)}.\no
\end{eqnarray}
Here each of $(i)$ and $(ii)$ is a sum of O$(d^2)$ i.i.d. Bernoulli($q$)
random variables minus their expectations. Therefore similar to Lemma
\ref{lem:d-sqrtd}, both $(i)$ and $(ii)$ can be shown to be O($\sqrt{12qd^2\log
n}$) with a probability at least $1-o(1/m^4)$. For $(iii)$ we can say
\[\sum_{u'\in N_G(u)\setminus
N_{G_p}(u)}\big(d_u'-1-q(d-1)\big)\leq d_u\max_{u'\in N_G(u)\setminus
N_{G_p}(u)}(|d_{u'}-1-q(d-1)|).\]
Now using Lemma \ref{lem:d-sqrtd}
for $d_u$ and each term $d_{u'}-1-q(d-1)$ we can say (iii) is $O\left(~[dq+\sqrt{12qd\log
n}~]\sqrt{12qd\log n}~\right)$ with a probability at least
$1-o(d/m^4)$. These finish the proof of Lemma \ref{lem:d-sqrtd-2}.
%Need to Define: a picture is helpful here
\enp\end{proof}
The final step in bounding the martingale difference is to apply
Lemmas \ref{lem:d-sqrtd}, \ref{lem:d-sqrtd-2} and the union bound to event
$L=\bigcap_{r=0}^{m_2}\bigcap_{u=1}^{~n}(L_u\cap K_u)$ and obtain
$\mb{P}(L^c)=o(1/m^2)$.

Hence for the martingale difference we have
$$ |Z_i-Z_{i-1}|\mbf{1}_{L}\leq O(dq+dq^{1/2}+(dq)^{3/2})\log n.$$
Note that Azuma's inequality cannot be used directly, since the martingale difference $|Z_i-Z_{i-1}|$ can be large
outside the set $L$. But the complement of $L$ has very low probability and we can use the following variation of Azuma's
inequality.
\begin{proposition}[Kim \cite{Kim}] Consider a martingale
$\{Y_i\}_{i=0}^n$ adaptive to a filtration $\{\mc{B}_i\}_{i=0}^n$.
If for all $k$ there are  $A_{k-1}\in\mc{B}_{k-1}$ such that
$\mathbf{E}[e^{\om Y_k}|\mc{B}_{k-1}]\mbf{1}_{A_{k-1}}\leq C_k$ for all
$k=1,2,\cdots,n$ with $C_k\geq1$ for all $k$, then
\[\mathbf{P}(Y-\mathbf{E}[Y]\geq \la)\leq e^{-\la\om}\prod_{k=1}^nC_k+\mathbf{P}(\cup_{k=0}^{n-1}A_k)\]
\end{proposition}
\begin{proof}[of Theorem \ref{thm:main-regular}]
Applying the above proposition for a large enough constant $c'>0$
gives
\begin{equation*}
\mb{P}\left(|X_r| > c'\sqrt{6r\log^3 n
\left(dq+d(q)^{1/2}+(dq)^{3/2}\right)^2} \right)\leq e^{-3\log
n}+\mb{P}(L^c)=o(\f{1}{m^2}).
\end{equation*}
Now using the fact that $G_p$ has $r$ edges with probability at least $1/n$, the same event in the random model $G_{\mc{N}_r}$ has probability $o(1/m)$.
A similar bound holds for $Y_r$ since the martingale
difference for $Y_r$ is
$O(|dq(d_u-qd)|)=O((dq)^{3/2}\log^{1/2}n))$ using Lemma
\ref{lem:d-sqrtd}.

Therefore defining $\hat{T}_r^{(2)}=
c'(dq+d(q)^{1/2}+(dq)^{3/2})\sqrt{6r\log^3 n}$ , we only need to show
\[
\sum_{r=0}^{m_2}\f{(dq+d(q)^{1/2}+(dq)^{3/2})\sqrt{6r\log^3
n}}{(2m-2r)^2}=o(1).\]

But using $ndq=2m-2r$ we have
\begin{eqnarray}
&&\sum_{r=0}^{m_2}\f{\left(dq+dq^{1/2}+(dq)^{3/2}\right)\sqrt{6r\log^3
n}}{n^2d^2q^2}\no\\
&=&\sum_{r=0}^{m_2}O\left(\f{d^{1/2}\log^{1.5}n}{n^{1/2}(2m-2r)}+\f{d\log^{1.5}n}{(2m-2r)^{3/2}}+\f{d^{1/2}\log^{1.5}n}{n(2m-2r)^{1/2}}\right)\no\\
&=&O\Big(\f{d^{1/2}\log(nd)}{n^{1/2}}+\f{d}{(n\log^3n
)^{1/2}}+\f{d}{n^{1/2}}\Big)\log^{1.5}n=o(1)\no
\end{eqnarray}
for $d=O(n^{1/2-\tau})$.
\enp\end{proof}
%-------------------------- PROBABILITY OF FAILURE -----------------------------------------
\section{Probability of Failure of Procedure $\mrm{A}$}\label{sec:prob-fail}

In this section we will prove Lemma \ref{lem:prob-fail} from Section \ref{sec:analysis}.
First we present the following remark.
\begin{remark}\label{rem:upperbound}
Lemma \ref{lem:main-general-conc} gives an
upper bound for the number of simple graphs with degree sequence
$\bar{d}$ independently from all known formulas for $|\mc{L}(\bar{d})|$.
If $d_{\max}=O(m^{1/4-\tau})$ then
\[|\mc{L}(\bar{d})|\leq
e^{-\la(\bar{d})-\la^2(\bar{d})+o(1)}\f{\prod_{r=0}^m{2m-2r\choose
2}}{m!\prod_{i=1}^nd_i!}.\]
In this section we will show that the above inequality is
in fact an equality. This is done by proving that the probability of failure of Procedure $\mrm{A}$
is very small.
\end{remark}
 First we will
characterize the degree sequence of the partial graph that is generated up
to the time of failure. Then we apply the upper bound of Remark \ref{rem:upperbound}
to derive an upper
bound on the probability of failure and show that it is $o(1)$.
\begin{lemma}\label{lem:fail-s-bound}
If Procedure $\mrm{A}$ fails in step $s$ then $2m-2s\leq
d_{\max}^2+1$.
\end{lemma}
\begin{proof}
Procedure $\mrm{A}$ fails when there is no suitable pair left to
choose. If the failure occurs in step $s$ then the number of unsuitable
edges is equal to the total number of possible pairs, that is ${2m-2s\choose2}$.
On the other hand, it can be easily shown that the number of unsuitable
edges at step $s$ is at most $d_{\max}^2(2m-2s)/2$ (see Corollary
3.1 in \cite{StegerWormald} for more detail). Therefore $2m-2s\leq
d_{\max}^2+1$.
\enp\end{proof}
Failure in step $s$ means there are some $W_i$'s which have
unmatched mini-vertices ($d_i^{(s)}\neq0$). Let us call them
``unfinished'' $W_i$'s. Since the algorithm fails, any two
unfinished $W_i$'s should be already connected. Hence there are at
most $d_{\max}$ of them. This is because for all $i$: $|W_i|=d_i\leq d_{\max}$.
The main goal is now to show that this scenario is a very rare event. Without loss of
generality assume that $W_1,W_2,\ldots,W_k$ are all the unfinished sets.
The argument given above shows $k\leq d_{\max}$. Moreover, by construction $k\leq
2m-2s$. The algorithm up to this step has created a partial matching
$\mc{M}_s$ where graph $G_{\mc{M}_s}$ is simple and has degree sequence
$\bar{d}^{(s)}=(d_1-d_1^{(s)},\ldots,d_k-d_k^{(s)},d_{k+1},\ldots,d_n)$.
Let $A_{d_1^{(s)},\ldots,d_k^{(s)}}$ denote the above event of
failure. Hence
\begin{equation}
\mb{P}(\textrm{fail})=\sum_{2m-2s=2}^{d_{\max}^2+1}\sum_{k=1}^{\max(d_{\max},2m-2s)}\sum_{i_1,\ldots,i_k=1}^n\mb{P}_{\mrm{A}}(A_{d_1^{(s)},\ldots,d_k^{(s)}}).
\label{eq:81}
\end{equation}
The following lemma is the central part of the proof.
\begin{lemma}\label{lem:fail-A-bound}
The probability of the event that Procedure $\mrm{A}$ fails in step $s$
and the vertices $v_1,\ldots,v_k$ are the only unfinished vertices; i.e.
$d_i^{(s)}\neq 0~i=1,\cdots,k$, is at most
\[\left(1+o(1)\right)\f{d_{\max}^{k(k-1)}\prod_{i=1}^kd_i^{d_i^{(s)}}}{m^{{k\choose2}}(2m)^{2m-2s}}{2m-2s\choose
d_1^{(s)},\ldots,d_k^{(s)}}.\]
\end{lemma}
\begin{proof}
Following the above notation, the event that we are considering is denoted by
$A_{d_1^{(s)},\ldots,d_k^{(s)}}$. Note that graph $G_{\mc{M}_s}$ should have a clique of size $k$ on vertices
$v_1,\ldots,v_k$. Therefore, the number of such graphs should be less
than $|\mc{L}(\bar{d}_k^{(s)})|$ where
$\bar{d}_k^{(s)}$=$(d_1-d_1^{(s)}-(k-1),\ldots,d_k-d_k^{(s)}-(k-1),d_{k+1},\ldots,d_n)$.
Thus, $\mb{P}_{\mrm{A}}(A_{d_1^{(s)},\ldots,d_k^{(s)}})$ is at most
$|\mc{L}(\bar{d}_k^{(s)})|~\mb{P}_{\mrm{A}}(G_{\mc{M}_s})$. On the
other hand we can use Remark 1 to derive an upper bound for
$|\mc{L}(\bar{d}_k^{(s)})|$ because $m-s$ and $k$ are small relative
to $m$ and it is easy to show that $d_{\max}=O([s-{k\choose2}]^{1/4-\tau})$. The result of these steps is
\[\mb{P}_{\mrm{A}}(A_{d_1^{(s)},\ldots,d_k^{(s)}})\leq
\left(\f{(2s-k(k-1))!~~~~\exp\left[-\la(\bar{d}_k^{(s)})-\la^2(\bar{d}_k^{(s)})+o(1)\right]}{[s-{k\choose2}]!~~~~2^{s-{k\choose2}}~~~~\prod_{i=1}^n(d_i^{(s)})!}\right)\mb{P}_{\mrm{A}}(G_{\mc{M}_s}).\]
The next step is to bound $\mb{P}_{\mrm{A}}(G_{\mc{M}_s})$. We can
use the same methodology as in the beginning of Section
\ref{sec:analysis} to derive
\begin{eqnarray}
\mb{P}_{\mrm{A}}(G_{\mc{M}_s})&=&\f{\prod_{i=1}^nd_i!}{\left(\prod_{i=1}^k\left[d_i^{(s)}\right]!\right)}\sum_{\mc{N}_s\in
S(\mc{M}_s)}\mb{P}_{\mrm{A}}(\mc{N}_s)\no\\
&=&~s!~\exp\left(-\f{\sum_{i\sim_{G_s}j}d_id_j}{4m}+o(1)\right)\prod_{r=0}^{s-1}\f{1}{{2m-2r\choose2}-\psi_r(G_{\mc{M}_s})}\no\\
&=&~s!~\exp\left(\f{m}{s}\la(\bar{d})+\f{m^2}{s^2}\la^2(\bar{d})+o(1)\right)\prod_{r=0}^{s-1}\f{1}{{2m-2r\choose2}}.\no
\end{eqnarray}
Similar to $\psi_r$, the quantity $\psi_r(G_{\mc{M}_s})$ is an approximation for the expected value of $\Psi_r$ conditioned on obtaining $G_{\mc{M}_s}$ at step $s$.
Now using the simple algebraic approximation
\begin{equation*}
\f{m}{s}\la(\bar{d})+\f{m^2}{s^2}\la^2(\bar{d})-\la(\bar{d}_k^{(s)})-\la^2(\bar{d}_k^{(s)})=O\left(\la(\bar{d})\left[\la(\bar{d})-\la(\bar{d}_k^{(s)})\right]\right)=O(\f{d_{\max}^4}{m^2})=o(1)
\end{equation*}
the following is true
\begin{eqnarray}
\mb{P}_{\mrm{A}}(A_{d_1^{(s)},\ldots,d_k^{(s)}})&\leq&e^{o(1)}\f{[2s-k(k-1)]!~(2m-2s)!~s!~2^{{k\choose2}}~\prod_{i=1}^kd_i!}{[s-{k\choose2}]!~(2m)!~\prod_{i=1}^k\big[(d_i^{(s)})!(d_i-k-d_i^{(s)}+1)!\big]}\no\\
&\leq&e^{o(1)}\f{\prod_{i=1}^kd_i^{d_i^{(s)}+k-1}}{\prod_{\ell=2s+1}^{2m}\ell~\prod_{j=1}^{k\choose2}(2s-2j+1)}{2m-2s\choose
d_1^{(s)},\ldots,d_k^{(s)}}.\label{eq:80}
\end{eqnarray}
The next step is to use $m-s=O(d_{\max}^2)$ and $k=O(d_{\max})$ to show that
$\prod_{j=1}^{k\choose2}(2s-2j+1)\geq m^{{k\choose2}}$ and
$(1/m^{2m-2s})\prod_{\ell=2s+1}^{2m}\ell\geq e^{-O(d_{\max}^4/m)}$.
These two facts combined with equation (\ref{eq:80}) finish the proof of Lemma
\ref{lem:fail-A-bound}.
\enp\end{proof}
Now we are ready to prove the main result of this section.
\begin{proof}[of Lemma \ref{lem:prob-fail}]
First, we show that the event of failure has a negligible probability
when there is only one unfinished vertex left, i.e., when $k=1$. In this case Lemma
\ref{lem:fail-A-bound} simplifies to
$\mb{P}_{\mrm{A}}(A_{d_1^{(s)}})=O\left((\f{D}{m})^{2m-2s}\right)$.
Therefore, summing over all possibilities of $k=1$ gives
\[\sum_{2m-2s=2}^{d_{\max}^2+1}\sum_{i=1}^n\mb{P}_{\mrm{A}}(A_{d_i^{(s)}})=
O\left(\sum_{2m-2s=2}^{d_{\max}^2+1}\f{d_{\max}^{2m-2s-1}}{m^{2m-2s-1}}\right)=O(\f{d_{\max}}{m})=o(1).\]
For $k>1$ we use Lemma \ref{lem:fail-A-bound} differently. Using
$d_{\max}^{k(k-1)}/m^{{k\choose2}}\leq d_{\max}^2/m$ and equation (\ref{eq:81}) we
have
\begin{multline*}
\mb{P}(\textrm{fail}) \leq o(1) \\
+\f{e^{o(1)}d_{\max}^2}{m}\sum_{2m-2s=2}^{d_{\max}^2+1}\f{\overbrace{\sum_{k=2}^{\max(d_{\max},2m-2s)}\sum_{i_1,\ldots,i_k=1}^n\prod_{i=1}^kd_i^{d_i^{(s)}}{2m-2s\choose
d_1^{(s)},\ldots,d_k^{(s)}}}^{(a)}}{(2m)^{2m-2s}}.
\end{multline*}
Now note that the double sum (a) is at most
$(d_1+\ldots+d_n)^{2m-2s}=(2m)^{2m-2s}$ since
$\sum_{i=1}^kd_i^{(s)}=2m-2s$. Therefore
\[\mb{P}(\textrm{fail})\leq o(1)+e^{o(1)}
\f{d_{\max}^2}{m}\sum_{2m-2s=2}^{d_{\max}^2+1}1=O(\f{d_{\max}^4}{m})=o(1).\]
\enp\end{proof}

%---------------------------------------------------------------------- RUNNING TIME
\section{Running Time of Procedure $\mrm{A}$}\label{sec:procB-run-time}

In this section we prove Lemma \ref{lem:rtime}.
\begin{proof}
Our proof is very similar to the analysis of Steger and Wormald \cite{StegerWormald}.
They use a non-trivial data structure and algorithm to efficiently choose a pair of vertices $v_i\in V$ and $v_j\in V$ with probabilities proportional to
$\hat{d}_i$ and $ \hat{d_j}$ respectively. They explain their methods for regular graphs but they only use the fact that
the maximum degree is bounded.  We include their analysis in Section \ref{subsec:gen_sw} for the sake of completeness.

We need to add a few steps to their method.  After choosing vertices $v_i$ and $v_j$ with the above probabilities, toss a biased coin that comes head with probability
$1-d_id_j/4m$.  Accept the pair $\{v_i,v_j\} $ if the coin shows head, $i\neq j$, and $\{v_i,v_j\}\notin
E$. Add $\{v_i,v_j\}$ to $E$ and reduce each of $\hat{d}_i,~\hat{d}_j$
by $1$. Otherwise reject the pair $\{v_i,v_j\}$ and repeat. The expected number of repeats is bounded by a constant because
$d_{\max}=O(m^{1/4-\tau})$ and therefore $1-d_id_j/4m>1/2$.

Efficient calculation of $P$ is also straightforward.  Note that
\[
p_{ij}=\f{(1-d_id_j/4m)d_i^{(r)}d_j^{(r)}}{{2m-2r\choose2}-\Psi_r(\mc{N})}.
\]
Therefore, $p_{ij}$ can be easily calculated from ${2m-2r\choose2}-\Psi_r(\mc{N})$.
At the beginning of Procedure $\mrm{A}$ we have
\[{2m\choose2}-\Psi({\mc{N}_0}) = {2m\choose2}-\sum_{u}{d_u\choose2}-\frac{(\sum_ud_u^2)^2-\sum_ud_u^4}{8m}
\]
which can be calculated with $O(n)$ operations. Now we show that in step $r+1$, $p_{ij}$ can be updated from step $r$ with $O(d_{\max})$ operations. This is because by choosing a pair $\{v_i,v_j\}$ at step $r+1$:
\begin{multline*}
\left[{2m-2r-2\choose2}-\Psi_{r+1}(\mc{N})\right]-\left[{2m-2r\choose2}-\Psi_r(\mc{N})\right]
 \\
 = \sum_{(v_{a},v_b)\in E_{r+1}}  d_a^{(r+1)}d_b^{(r+1)}(1-\frac{d_ad_b}{4m})- \sum_{(v_{a},v_b)\in E_{r}}  d_a^{(r)}d_b^{(r)}(1-\frac{d_ad_b}{4m})\\
 =-d_i^{(r)}d_j^{(r)}(1-\frac{d_id_j}{4m}) - \sum_{(v_{i'},v_i)\in E_r} d_{i'}^{(r)}(1-\frac{d_id_{i'}}{4m}) - \sum_{(v_{j'},v_j)\in E_r} d_{j'}^{(r)}(1-\frac{d_jd_{j'}}{4m})\\
=-d_i^{(r)}d_j^{(r)}(1-\frac{d_id_j}{4m}) + \Xi_{i,r} + \Xi_{j,r} + \frac{d_i+d_j}{4m}\Omega_r + O_{i,r} + O_{j,r}
  % - \sum_{v_{j'}\sim_{G_{{\mc{N}_r}}}v_j} d_{j'}^{(r)}(1-\frac{d_jd_{j'}}{4m})
\end{multline*}
where $\Xi_{i,r}=\sum_{v_{i'}\sim_{G_{{\mc{N}_r}}}v_i} d_{i'}^{(r)}(1-\frac{d_id_{i'}}{4m})$, $\Omega_r=\sum_{i'=1}^nd_{i'}^rd_{i'}$, and $O_{i,r}=d_i^{(r)}(1-d_i^2/4m)-(2m-2r)$.
It is clear to see from $\Omega_{r+1}-\Omega_r=-d_i-d_j$ that $\Omega_r$ can be updated at each step by only one operation, and the calculation of $O_{i,r}\,,~O_{j,r}$ takes constant time. Moreover, each of $\Xi_{i,r}, \Xi_{j,r}$ is a summation with at most $d_{\max}$ terms.  We will show in the next section that { it is possible to} find
neighbors of $v_i$ and $v_j$ in $G_{{\mc{N}_r}}$ with $O(d_{\max})$ operations. Therefore
$\Xi_{i,r}, \Xi_{j,r}$ can be calculated with $O(d_{\max})$ operations.
Thus the running time of the new implementation of
Procedure $\mrm{A}$ is O($m\,d_{\max}$) for general degree
sequences.  Now using Lemma \ref{lem:prob-fail}, the running
time of Procedure $\mrm{A}$ is of O($m\,d_{\max}$).
\enp\end{proof}

\subsection{Steger and Wormald's method for choosing a suitable pair}
\label{subsec:gen_sw}
Steger and Wormald's (SW) \cite{StegerWormald} implementation has three phases and uses the
configuration model.

In the first phase, the algorithm puts all of the mini-vertices in an array $L$ where all of the
matched mini-vertices are kept in the front.  It is also assumed
that the members of each pair of matched mini-vertices will be two
consecutive elements of $L$. There is another array $I$ that keeps
location of each mini-vertex inside array $L$. Then two elements of $L$ (selected uniformly at
random) can be checked for suitability in
time $O(d_{\max})$. This is because from $I$ we can find the neighbors
of the selected elements in the partially constructed graph $G_{\mc{N}_r}$.
Note that in our modification (Procedure $\mrm{A}$), the
pair is accepted with probability $1-d_id_j/4m$ when they belong to
$W_i,W_j$. This also completes the above argument
for updating $\Psi_r(\mc{N})$ with $O(d_{\max})$ operations. Repeat the
above till a suitable pair is found then update $L$ and $I$.

Phase 1 ends when the number of remaining mini-vertices falls below
$2d_{\max}^2$. Hence using Corollary 3.1 in \cite{StegerWormald},
throughout phase 1 the number of suitable pairs is more than half of the
total number of available pairs. Therefore, the expected number of
repetitions in the above process is at most $2$. This means the
expected running time of phase 1 is O($m\,d_{\max}$).

Phase 2 starts when the number of available mini-vertices is less than
$2d_{\max}^2$ and finishes when the number of available vertices is at least
$2d_{\max}$.  In this phase instead of choosing the mini-vertices,
choose a pair of vertices of $G_{\mc{N}_r}$ (two random set $W_i,W_j$ in the
configuration model) from the set of vertices that are not fully
matched. Repeat the above till $v_i,v_j$ is not already an edge in $G_{\mc{N}_r}$.
Again the expected number of repetitions is at most 2. Now randomly
choose one mini-vertex in each selected $W_i$. If both of the
mini-vertices are not matched yet add the edge, otherwise pick
another two mini-vertices.  The expected number of repetitions here is
at most O($d_{\max}^2$) and hence the expected running time of the phase
2 is at most O($d_{\max}^4$).

Phase 3 starts when the number of available vertices (not fully matched $W_i$'s) is less
than $2d_{\max}$.  We can construct a graph $H$, in time
O($d_{\max}^2$), that indicates the set of all possible connections.
Now choose an edge $\{v_i,v_j\}$ of $H$ uniformly at random and accept it
with probability $\hat{d}_i\hat{d}_j/d_{\max}^2$. Again, the expected
number of repetitions will be at most O($d_{\max}^2$).  Update $H$
in constant time and repeat the above till $H$ is empty. Therefore
the expected running time of phase 3 is also O($d_{\max}^4$).

Hence, the total running time for $d_{\max}=O(m^{1/4-\tau})$ will be
O($m\,d_{\max}$).

%------------------------------------------------ Generalizing Kim-Vu
\section{Generalizing Kim and Vu's Analysis}\label{sec:Kim-Vu-Calc}

The aim of this section is to show equation (\ref{eq:kimvuconc}) via generalization of Kim and Vu's analysis \cite{Kim-Vu}.
Let us define
\[
f(\mc{N})=\prod_{r=0}^{m-1}\f{{2m-2r\choose2}-\psi_r(G)}{{2m-2r\choose2}-\Psi_r(\mc{N})}
\]
then equation (\ref{eq:kimvuconc}) is equivalent to
\begin{equation}
\mb{E}(f(\mc{N})) = 1+ o(1) \label{eq:Ef=1}
\end{equation}
where the expectation is with respect to the uniform distribution on the set $S(\mc{M})$ of all $m!$ orderings of the matching $\mc{M}$. Proof of equation (\ref{eq:Ef=1}) is done by partitioning the set $S(\mc{M})$ into smaller subsets and looking at the deviation of $f$ on each set separately. The partition is explained in Section \ref{subsec:partitions}. But before that we need to define some notation.

\subsection{Definitions}\label{subsec:kim-vu-defs}

In Section \ref{sec:analysis} we saw that the probability of choosing an
edge between $W_i$ and $W_j$ at step $r+1$ of Procedure $\mrm{A}$ is equal
to
$(1-\f{d_id_j}{4m})\left[{2m-2r\choose2}-\Psi_r(\mc{N})\right]^{-1}$
where
$$\Psi_r(\mc{N})=\sum_{\{v_i,v_j\}\notin E_r}d_i^{(r)}d_j^{(r)}+\sum_{\{v_i,v_j\}~\in~E_r}d_i^{(r)}d_j^{(r)}\f{d_id_j}{4m}.$$
To simplify the notation, throughout the rest of this section, we will use $\Psi_r$ and $\Delta_r$ to denote $\Psi_r(\mc{N})$ and $\Delta_r(\mc{N})$ respectively.
We will also use the notation $\{v_i,v_j\}$ and $\{i,j\}$ interchangeably. Moreover, the notation $\{i,j\}$ includes the cases of $i=j$ as well.

For our analysis we need to write $\Psi_r=\Delta_r+\Lambda_r$ where
\begin{eqnarray}
\Delta_r&=&{2m-2r\choose2}-\sum_{\{i,j\}~\in~E_r}d_i^{(r)}d_j^{(r)},\no\\
\Lambda_r&=&\sum_{\stackrel{\{i,j\}}{i\neq j}}d_i^{(r)}d_j^{(r)}\f{d_id_j}{4m}-\sum_{\stackrel{\{i,j\}~\notin~E_r}{i\neq j}}d_i^{(r)}d_j^{(r)}\f{d_id_j}{4m}.\no
\end{eqnarray}
Notice that $\Delta_r$ counts the number of possibilities for creating
a self loop ($i=j$) or making double edges. We distinguish between these two cases by an extra index. That is
\begin{eqnarray}
\Delta_r^{(1)}&=&\sum_{i=1}^n{d_i^{(r)}\choose2}=\textrm{\# of self loops, and}\no\\
\Delta_r^{(2)}&=&\Delta_r-\Delta_r^{(1)}=\textrm{\# of double edges.}\no
\end{eqnarray}
Note that since all the existing pairs are suitable, the only type of multiple pairs that can be created at step $r+1$ are double pairs.
Moreover,
\begin{eqnarray}
4m\Lambda_r&=&\sum_{\stackrel{\{i,j\}}{i\neq j}}d_i^{(r)}d_j^{(r)}d_id_j-\sum_{\stackrel{\{i,j\}~\notin~E_r}{i\neq j}}d_i^{(r)}d_j^{(r)}d_id_j\no\\
&=&\f{(\sum_{i=1}^nd_i^{(r)}d_i)^2-\sum_{i=1}^n(d_i^{(r)})^2d_i^2}{2}-\sum_{\stackrel{\{i,j\}~\notin~E_r}{i\neq j}}d_i^{(r)}d_j^{(r)}d_id_j.\no
\end{eqnarray}
We distinguish between these three summations by adding a numerical index to
$\Lambda_r$; i.e.
$$
\Lambda_r^{(1)}=\sum_{i=1}^nd_i^{(r)}d_i~~~,~~~\Lambda_r^{(2)}=\sum_{i=1}^n(d_i^{(r)})^2d_i^2~~~~,~~~~\Lambda_r^{(3)}=\sum_{\stackrel{\{i,j\}~\notin~E_r}{i\neq
j}}d_i^{(r)}d_j^{(r)}d_id_j.
$$
Hence,
$$
\Lambda_r =
\f{(\Lambda_r^{(1)})^2-\Lambda_r^{(2)}}{8m}-\f{\Lambda_r^{(3)}}{4m}.
$$
% --------------------------------------------------------------------- Bounds on Deltas
The following simple bounds will be very useful throughout Section
\ref{sec:Kim-Vu-Calc}.
\begin{lemma}\label{lem:del-lam-bounds} For all $r$ the following equations
hold.
\begin{itemize}
\item[(i)] $\Delta_r\leq\f{(2m-2r)d_{\max}^2}{2}$
\item[(ii)] $\Lambda_r^{(1)}\leq d_{\max}(2m-2r)$
\item[(iii)] $\Lambda_r\leq \f{(2m-2r)^2d_{\max}^2}{8m}$
\end{itemize}
\end{lemma}
\begin{proof}
\begin{itemize}
\item[(i)] At step $r$ there are $2m-2r$ mini-vertices left and for each each $u\in W_i$
there are at most $d_{\max}-1$ mini-vertices in $W_i$ that $u$ can
connect to. Hence, $\Delta_r^{(1)}\leq\f{(2m-2r)(d_{\max}-1)}{2}$.
Similarly $u$ can connect to at most $(d_{\max}-1)^2$ mini-vertices in some
$W_j$ with $i\neq j$ to create a double edge. Thus, $\Delta_r^{(2)}\leq\f{(2m-2r)(d_{\max}-1)^2}{2}$. Now
using $\Delta_r=\Delta_r^{(1)}+\Delta_r^{(2)}$ the proof of (i) is clear.

\item[(ii)] $\Lambda_r^{(1)}\leq d_{\max}\sum_ud_u^{(r)}=d_{\max}(2m-2r)$.

\item[(iii)] It follows from the definition of $\Lambda_r$ that
$$\Lambda_r=\sum_{\{i,j\}~\in~E_r}d_i^{(r)}d_j^{(r)}\f{d_id_j}{4m}\leq\f{d_{\max}^2}{4m}\sum_{\{i,j\}~\in~E_r}d_i^{(r)}d_j^{(r)}\leq\f{d_{\max}^2}{4m}{2m-2r\choose2}.$$
\end{itemize}
\enp\end{proof}
In order to define $\psi_r$ we look at a slightly similar model. Recall that $G_{\mc{N}_r}$ is the partial graph that is constructed up to step $r$.
Imposing the uniform distribution on $S(\mc{M})$, graph $G_{\mc{N}_r}$ turns to a random subgraph of $G$ that has exactly $r$ edges. { We } can approximate this graph by a different random subgraph of $G$. This is done, by selecting each edge of $G$ independently with probability $p_r=r/m$ and denoting the resulted graph by $G_{p_r}$. Now using $G_{p_r}$ as an approximation to $G_{\mc{N}_r}$, we are ready evaluate quantities $\mb{E}_{p_r}(\Delta_r^{(1)})$, $\mb{E}_{p_r}(\Delta_r^{(2)})$, $\mb{E}_{p_r}(\Lambda_r^{(1)})$,
$\mb{E}_{p_r}(\Lambda_r^{(2)})$, and $\mb{E}_{p_r}(\Lambda_r^{(3)})$.
Throughout this section we often use the notations $\Delta_{p_r}^{(i)}$, $\Lambda_{p_r}^{(i)}$, and $\Psi_{p_r}$ to emphasis that the model is $G_{p_r}$ instead of $G_{\mc{N}_r}$.
%----------------------------------------------------------------------------- AVERAGES
\begin{lemma} \label{lem:averages} For each $r$ the following equations hold.
\begin{itemize}
\item[(i)]$\mb{E}_{p_r}(\Delta_r^{(1)})=\f{(2m-2r)^2}{2}\left(\f{\sum_{i=1}^n{d_i\choose2}}{2m^2}\right)=\f{(2m-2r)^2}{2}\big(\f{\la(\bar{d})}{m}\big)$

\item[(ii)]$\mb{E}_{p_r}(\Delta_r^{(2)})=\f{(2m-2r)^2}{2}\left(\f{r\sum_{i\sim_Gj}(d_i-1)(d_j-1)}{2m^3}\right)$

\item[(iii)]$\mb{E}_{p_r}(\Lambda_r^{(1)})=(2m-2r)\f{\sum_{i=1}^nd_i^2}{2m}$

\item[(iv)]$\mb{E}_{p_r}(\Lambda_r^{(2)})=(2m-2r)^2\f{\sum_{i=1}^nd_i^4}{4m^2}+2r(2m-2r)\f{\sum_{i=1}^nd_i^3}{4m^2}$

\item[(v)]$\mb{E}_{p_r}(\Lambda_r^{(3)})=\f{(2m-2r)^2}{2}\left(\f{r\sum_{i\sim_Gj}d_id_j(d_i-1)(d_j-1)}{2m^3}\right)$
\end{itemize}
\end{lemma}
\begin{proof}
\begin{itemize}
\item[(i)]
In the random model of $G_{p_r}$ each edge has a
probability of $\f{r}{m}$ to be chosen. Let $X_i$ the number of unsuitable edges that
connect two mini-vertices of $W_i$ at $(r+1)^{th}$ step of creating
$\mc{N}$. Hence, $X_i$ is equal to the number of unordered tuples $\{j,i,k\}$ where $\{j,i\},\{i,k\}\in E(G)\stm
E(G_{\mc{N}_r})$ which gives
\begin{eqnarray}
\Delta_r^{(1)}&=&\sum_{i=1}^nX_i\label{eq:delta1=sumXi}.
\end{eqnarray}
On the other hand for a fixed $i$, the number of tuples $\{j,i,k\}$
where $\{j,i\},\{i,k\}\in E(G)$ is exactly ${d_i\choose2}$, and with
probability $(1-\f{r}{m})^2$ the edges $\{j,i\},\{i,k\}$ do not belong to $E(G_{\mc{N}_r})$. Thus,
the equality $\mb{E}(X_i)=(1-\f{r}{m})^2{d_i\choose2}$ holds and it can be used in (\ref{eq:delta1=sumXi}) to complete the proof of (i).

\item[(ii)]
Define $Y_{ij}$ to be the number of unsuitable
edges between $W_i$ and $W_j$ at $(r+1)^{th}$ step of creating $\mc{N}$.
It is not hard to see that $Y_{ij}$ also counts the number of unordered tuples $\{k,i,j,l\}$ where
$\{i,j\}\in E(G_{\mc{N}_r})$ but $\{k,i\},\{j,l\}\in E(G)\stm
E(G_{\mc{N}_r})$. Hence,
\begin{eqnarray}
\Delta_r^{(2)}&=&\sum_{i\sim_Gj}Y_{ij}.\label{eq:delta2=sumYij}
\end{eqnarray}
On the other hand for a fixed $i\sim_Gj$, the number of tuples
$\{k,i,j,l\}$ where $\{k,i\},\{j,l\}$ belong to $E(G)$ is exactly
$(d_i-1)(d_j-1)$. Moreover, the edges $\{k,i\},\{j,l\}$ do not belong to $E(G_{\mc{N}_r})$ with probability $(1-\f{r}{m})^2$,
and the edge $\{i,j\}$ belongs to $E(G_{\mc{N}_r})$ with probability $\f{r}{m}$. This gives the equality
$\mb{E}(Y_{ij})=\f{r}{m}(1-\f{r}{m})^2(d_i-1)(d_j-1)$ which can be used with (\ref{eq:delta2=sumYij}) to complete the proof of $(ii)$.

\item[(iii)] The proof directly follows from $\mb{E}(d_i^{(r)})=(1-\f{r}{m})d_i$.

\item[(iv)] Since each $d_i^{(r)}$ is a summation of $d_i$ Bernoulli iid random variables, { We} can show
$$\mb{E}\bigg[(d_i^{(r)})^2\bigg]= (1-\f{r}{m})^2d_i^2 + \f{r}{m}(1-\f{r}{m})d_i$$
which proves $(iv)$.
\item[(v)] The proof is similar to (ii), except we are using the following instead of (\ref{eq:delta2=sumYij})
\begin{eqnarray}
\Lambda_r^{(3)}&=&\sum_{i\sim_Gj}\f{d_id_j}{4m}Y_{ij}.\no
\end{eqnarray}
\end{itemize}
\enp
\end{proof}
The next step is to define $\psi_r$ as an approximation to $\mb{E}_{p_r}(\Psi_r)$. For that we will use Lemma \ref{lem:averages} and the following two estimates
\begin{eqnarray}
\mb{E}_{p_r}(\frac{\Lambda_r^{(2)}}{8m})&=&\frac{(2m-2r)^2}{2}\left[O(\frac{d_{\max}^3}{m^2})+O(\frac{d_{\max}^2}{m^2}\frac{2r}{2m-2r})\right],\no\\
\mb{E}_{p_r}(\frac{\Lambda_r^{(3)}}{4m})&=&\frac{r(2m-2r)^2}{2}O(\frac{d_{\max}^4}{m^3}).\no
\end{eqnarray}
Note that here we used the bound
$$\sum_{i=1}^nd_i^s=\sum_{i\sim_G j}(d_i^{s-1}+d_j^{s-1})=O(md_{\max}^{s-1})$$
that will be repeatedly used in this section.

Now $\mb{E}_{p_r}(\Psi_r)$ is given by the following expression
\begin{multline}
\label{eq:av(Psi)}
\mb{E}_{p_r}(\Psi_r)=\frac{(2m-2r)^2}{2}\Bigg[\f{\la(\bar{d})}{m}+\f{r\sum_{i\sim_Gj}(d_i-1)(d_j-1)}{2m^3}\\
+\f{(\sum_{i=1}^nd_i^2)^2}{16m^3}+O\bigg(\frac{rd_{\max}^4}{m^3}+\frac{rd_{\max}^2}{(m-r)m^2}\bigg)\Bigg].
\end{multline}
\begin{definition} The expected value of $\Psi_r$ is denoted by $\psi_r$. i.e.  $\psi_r=\mb{E}_{p_r}(\Psi_r)$.
\end{definition}
The following lemma is equivalent to \eqref{eq:av(Psi)}.
\begin{lemma} For all $r$,
$$\psi_r=\frac{(2m-2r)^2}{2}\left(\f{\la(\bar{d})}{m}+\f{r\sum_{i\sim_Gj}(d_i-1)(d_j-1)}{2m^3}+\f{(\sum_{i=1}^nd_i^2)^2}{16m^3}+\varsigma_r\right)$$ where $\varsigma_r=O(\frac{rd_{\max}^4}{m^3}+\frac{rd_{\max}^2}{(m-r)m^2})$.
\end{lemma}
It is also straightforward to show that the following upper bound holds for $\psi_r$.
\begin{lemma}\label{lem:bound-on-psi}
For all $r$ the quantity $\psi_r$ is bounded above by  $O(\f{d_{\max}^2(2m-2r)^2}{2m})$.
\end{lemma}
Now we are ready to prove equation (\ref{eq:alg}).
%--------------------------------------------------
\subsection{Algebraic Proof of the Equation (\ref{eq:alg})}\label{subsec:alg-proof}
For simplicity, we define $\chi_G$ to be $\sum_{i\sim_Gj}(d_i-1)(d_j-1)$. Therefore,
{%\small
\begin{eqnarray}
&&\prod_{r=0}^{m-1}\f{{2m-2r\choose 2}}{{2m-2r\choose 2}-\psi_r}\no\\
&=&\prod_{r=0}^{m-1}\Big(1+\f{\psi_r}{{2m-2r\choose
2}-\psi_r}\Big)\no\\
&=&\prod_{r=0}^{m-1}\Big(1+\f{\f{\la(\bar{d})}{m}+\f{r\sum_{i\sim_Gj}(d_i-1)(d_j-1)}{2m^3}+\f{(\sum_id_i^2)^2}{16m^3}+\varsigma_r}{1-\f{1}{2m-2r}-O(\f{d_{\max}^2}{m})}\Big)\no\\
&=&\exp{\Big[\sum_{r=0}^{m-1}\log\Big(1+\f{\f{\la(\bar{d})}{m}+\f{r\chi_G}{2m^3}+\f{(\sum_id_i^2)^2}{16m^3}+\varsigma_r}{1-\f{1}{2m-2r}-O(\f{d_{\max}^2}{m})}\Big)\Big]}\no\\
&=&\exp{\Big[\sum_{r=0}^{m-1}\log\Big(1+\f{\la(\bar{d})}{m}+\f{r\chi_G}{2m^3}+\f{(\sum_id_i^2)^2}{16m^3}+O(\frac{d_{\max}^4}{m^2}+\frac{rd_{\max}^2}{(m-r)m^2})\Big)\Big]}\no\\
&=&\exp{\Big[\sum_{r=0}^{m-1}\Big(\f{\la(\bar{d})}{m}+\f{r\chi_G}{2m^3}+\f{(\sum_id_i^2)^2}{16m^3}+O(\frac{d_{\max}^4}{m^2}+\frac{rd_{\max}^2}{(m-r)m^2})\Big)\Big]}\label{eq:sixteen}\\
&=&\exp{\Big[\la(\bar{d})+\f{m(m-1)\chi_G}{4m^3}+\f{(\sum_id_i^2)^2}{16m^2}+O\big(\f{d_{\max}^4}{m}+\f{d_{\max}^2}{m}\log(2m)\big)\Big]}\no\\
&=&\exp\Big[\la(\bar{d})+\f{\chi_G}{4m}+\f{(\sum_id_i^2)^2}{16m^2}+o(1)\Big]\no\\
&=&\exp\Big[\la(\bar{d})+\f{\sum_{i\sim_Gj}d_id_j}{4m}-\f{\sum_{i\sim_Gj}(d_i+d_j)}{4m}+\f{1}{4}+\f{(\sum_id_i^2)^2}{16m^2}+o(1)\Big]\label{eq:seventeen1}\\
&=&\big(1+o(1)\big)\exp\Big[\la(\bar{d})+\la^2(\bar{d})+\f{\sum_{i\sim_Gj}d_id_j}{4m}\Big]\label{eq:seventeen3}
\end{eqnarray}
}
where (\ref{eq:sixteen}) uses $\log(1+x)=x-O(x^2)$ and
(\ref{eq:seventeen1}) uses $d_{\max}=O(m^{1/4-\tau})$. The bound $\f{\psi_r}{(2m-2r)^2}=O(\f{d_{\max}^2}{m})$ was used a few times as well.
\enp

%------------------------------------- SUBSECTION SETUP AND PARTITIONS
\subsection{Partitioning the set of orderings $S(\mc{M})$}\label{subsec:partitions}
In order to prove equation \eqref{eq:Ef=1}, we need to study the large deviation behavior of function $f$ on the set $S(\mc{M})$.
For that we partition the set $S(\mc{M})$ in four ``major'' steps. At each step, one subset
of $S(\mc{M})$ will be removed from it.
\begin{itemize}
\item[Step 1)] Consider those orderings $\mc{N}\in S(\mc{M})$ where at any state during the algorithm, the
number of unsuitable edges does not exceed a constant (strictly less than 1) fraction of the number of all
available edges. More specifically, for a small number $0<\tau\leq1/3$ let
$$S^*(\mc{M})=\left\{\mc{N}\in S(\mc{M})~|~\Psi_r(\mc{N})\leq(1-\tau/4){2m-2r\choose2}:~\forall~0\leq r\leq m-1\right\}.$$
Then the first element of the partition will be $S(\mc{M})\stm
S^*(\mc{M})$.

\item[Step 2)] Consider those orderings $\mc{N}$ from the set $S^*(\mc{M})$ for which $\Psi_r(\mc{N})-\psi_r>
T_r(\log n)^{1+\delta}$ for all $0\leq r\leq m-1$. The function $T_r$ will be defined in Section \ref{subsec:rg-KimVu-more-notation} and $\delta$ is a small positive constant. For example
$\delta<0.1$ works. Denote the set of all such $\mc{N}$ by $\mc{A}$.

\item[Step 3)] From the set $S^*(\mc{M})\stm\mc{A}$, remove those elements with $\Psi_r(\mc{N})>0$ for some $r$ with $2m-2r\leq (\log n)^{1+2\delta}$.
Put these elements in the set $\mc{B}$.

\item[Step 4)] The last element of the partition is the remaining subset $\mc{C}=S^*(\mc{M})\stm(\mc{A}\cup\mc{B})$.
\end{itemize}
The journey towards proving equation (\ref{eq:kimvuconc}) is divided into these five parts
\begin{eqnarray}
\mb{E}(f(\mc{N})\mrm{1}_{\mc{A}})&=&o(1)\label{eq:av-f1A=o1},\\
\mb{E}(f(\mc{N})\mrm{1}_{\mc{B}})&=&o(1)\label{eq:av-f1B=o1},\\
\mb{E}(f(\mc{N})\mrm{1}_{\mc{C}})&\leq&1+o(1)\label{eq:av-f1C<1+o1},\\
\mb{E}(f(\mc{N})\mrm{1}_{\mc{C}})&\geq&1-o(1)\label{eq:av-f1C>1-o1},\\
\mb{E}(f(\mc{N})\mrm{1}_{S(\mc{M})\stm
S^*(\mc{M})})&=&o(1).\label{eq:av-f1S*c=o1}
\end{eqnarray}
These parts will be all proved is Section \ref{subsec:four-lem}.
The hardest of these proofs is for \eqref{eq:av-f1A=o1} which
is carried out by partitioning the set $\mc{A}$ into further subsets and using Vu's inequality on them. The remaining proofs for (\ref{eq:av-f1B=o1})-(\ref{eq:av-f1S*c=o1}) are based on the standard combinatorial and algebraic bounds.

\subsection{More notation}\label{subsec:rg-KimVu-more-notation}
% ----------------------------------   Defining T(\lambda) and more partitioning
In order to prove (\ref{eq:av-f1A=o1}) we need more notation.
Remember from Section \ref{subsec:partitions} that $\delta>0$ is a very small
constant. Let $\om=(\log n)^\delta$. Let $\la_0=\om\log n$ and $\la_i=2^i\la_0$ for
$i=1,2,\ldots,L$. $L$ is such that $\la_{L-1}<cd_{\max}^2\log n\leq\la_L$
where $c$ is a large constant that is specified later.
\begin{definition}\label{def:T}
Let $q_r=(1-r/m),~p_r=1-q_r~\forall 0\leq r\leq m-1$. Then let
\begin{eqnarray}
\beta_r(\la)&=&c\sqrt{\la(m\,d_{\max}^2q_r^2+\la^2)(d_{\max}^2q_r+\la)},\no\\
\gamma_r(\la)&=&c\sqrt{\la(m\,d_{\max}^2q_r^3+\la^3)(d_{\max}^2q_r^2+\la^2)},\no\\
\nu_r&=&8m\,d_{\max}^2q_r^3.\no
\end{eqnarray}
Now the function $T_r$ for all $0\leq r\leq m-1$ is defined by
$$
T_r(\la)=\left\{
\begin{array}{ll}
3\beta_r(\la)+2\min(\g_r(\la),\nu_r)&\textrm{ if }2m-2r\geq \om\la.\\
\la^2/\om&\textrm{~Otherwise.}
\end{array}\right.
$$
\end{definition}
The intuition behind this definition will become clear when we use Vu's concentration inequality in Section \ref{subsubsec:proo-A-lemmas-a}.
Note that inequalities $\al_r(\la)\leq \beta_r(\la)$ and $\zeta_r(\la)\leq \beta_r(\la)$ hold and we will use them in Section \ref{subsec:four-lem} to simplify the computations. Moreover, with the above definition,
since $\la_i=2\la_{i-1}$, the following relation holds between $T_r(\la_i),~T_r(\la_{i-1})$.
\begin{eqnarray}
T_r(\la_i)\leq 8T_r(\la_{i-1}).\label{eq:T-ineq}
\end{eqnarray}
Now we will subpartition $\mc{A}$ and $\mc{B}$. Define subsets $A_0\subseteq A_1\subseteq\ldots\subseteq
A_L\subseteq S^*(\mc{M})$ by
$$A_i=\{\mc{N}\in S^*(\mc{M})~|~\Psi_r(\mc{N})-\psi_r<T_r(\la_i),~\forall~0\leq r\leq m-1\}.$$
Moreover, define $A_\infty$ by $A_\infty=S^*(\mc{M})\stm\cup_{i=0}^LA_i$. Then we have
$$\mc{A}=A_\infty\cup \left(\cup_{i=1}^LA_i\stm A_{i-1}\right).$$
Since the main objective of partitioning $\mc{A}$ is to prove \eqref{eq:av-f1A=o1}, we are only interested in finding upper bounds for $f(\mc{N})=\prod_{r=0}^{m-1}\left(1+\f{\Psi_r(\mc{N})-\psi_r}
{{2m-2r\choose2}-\Psi_r(\mc{N})}\right)$. Therefore, the cases with $\Psi_r(\mc{N})\leq \psi_r$ are not troublesome.

Let $K$ be an integer such that $2^{K-1}<(\log n)^{2+\delta}+1\leq
2^K$. Next step is to consider a chain of subsets $B_0\subseteq B_1\subseteq\cdots\subseteq
B_K\subseteq A_0$ that are defined by
$$B_j=\{\mc{N}\in A_0~|~\Psi_r(\mc{N})<2^j,~\forall~ r\geq (2m-\om\la_0)/2\}.$$
It is not hard to see that the set $\mc{C}$ that was defined in step 4 in Section \ref{subsec:partitions} is equal to the set $B_0$. Note that $T_r$'s are chosen such that for all $r\geq (2m-\om\la_0)/2$ we have
$T_r(\la_0)=\la_0\log n$ and by Lemma \ref{lem:bound-on-psi}, for all $r\geq (2m-\om\la_0)/2$ we have
$\psi_r=o(1)$. Thus, for all such $r$ and all elements of $A_0$,
$$\Psi_r<\la_0\log n+\psi_r<2^K.$$
This shows that $A_0=\big(\cup_{j=0}^K B_j\big)\cup \mc{C}$ and also $\mc{B}=\cup_{j=1}^K B_j$.

\subsection{Proofs of (\ref{eq:av-f1A=o1}), (\ref{eq:av-f1B=o1})
and (\ref{eq:av-f1C<1+o1})}\label{subsec:four-lem}
In this section we will bound the expected value $\mb{E}(f(\mc{N}))$ on the sets $A_\infty$, $\mc{C}$, and on each of the sets of the form $A_i\setminus A_{i-1}$ and $B_j\setminus B_{j-1}$.
\begin{lemma}\label{lem:f1A} For all $1\leq i\leq L$,
\begin{itemize}
\item[(a)] $\mb{P}(A_i\stm A_{i-1})\leq e^{-\Om(\la_i)}.$
\item[(b)] For all $\mc{N}$ in $A_i\stm A_{i-1}$ we have $f(\mc{N})\leq e^{o(\la_i)}$.
\end{itemize}
\end{lemma}
\begin{lemma}\label{lem:f1Ainf} For a large enough constant $c$,
\begin{itemize}
\item[(a)]$\mb{P}(A_\infty)\leq e^{-cd_{\max}^2\log n}.$
\item[(b)] For all $\mc{N}$ in $A_\infty$ we have $f(\mc{N})\leq e^{4d_{\max}^2\log n}.$
\end{itemize}
\end{lemma}
\begin{lemma}\label{lem:f1B} For all $1\leq j\leq K$,
\begin{itemize}
\item[(a)]$\mb{P}(B_j\stm B_{j-1})\leq e^{-\Om(2^{j/2}\log n)}$
\item[(b)] For all $\mc{N}$ in $B_j\stm B_{j-1}$ we have $f(\mc{N})\leq e^{O(2^{3j/4})}$.
\end{itemize}
\end{lemma}
\begin{lemma}\label{lem:f1C}
For all $\mc{N}\in \mc{C}$ we have $f(\mc{N})\leq 1+o(1)$.
\end{lemma}
Now it is easy to see that equation (\ref{eq:av-f1A=o1}) follows from Lemmas
\ref{lem:f1A} and \ref{lem:f1Ainf}. Note that by the definition of $K$ we
have $2^{K/4}\ll\log n$ which gives $2^{3j/4}\ll2^{j/2}\log n$. Thus, { we} can
deduce (\ref{eq:av-f1B=o1}) from Lemma \ref{lem:f1B}. Finally,
(\ref{eq:av-f1C<1+o1}) is consequence of Lemma \ref{lem:f1C}.

Proof of Lemma \ref{lem:f1A} uses Vu's concentration inequality but
for the other three lemmas, typical algebraic and combinatorial bounds
are sufficient. Throughout the rest of this section we present a quick
introduction to Vu's concentration inequality. Then we prove the above lemmas.

%------------------------------------------------ Concentration Inequalities
\subsubsection{Vu's Concentration inequality.}\label{subsubsec:vu-conc-inq}
Proofs of Lemmas \ref{lem:f1A}(a) and \ref{lem:f1Ainf}(a) use a very
strong concentration inequality proved by Vu \cite{Vu} which is a
generalized version of an earlier result by Kim and Vu
\cite{Kim-Vu2000}. Consider independent random variables
$t_1,t_2,\ldots,t_n$ with arbitrary distribution in $[0,1]$. Let
$Y(t_1,t_2,\ldots,t_n)$ be a polynomial of degree $k$ and
coefficients in $(0,1]$. For any multi-set $A$ of elements
$t_1,t_2,\ldots,t_n$ let $\p_AY$ denote the partial derivative of $Y$
with respect to variables in $A$. For example if $Y=t_1+t_1^3t_2^2$
and $A=\{t_1,t_1\},~B=\{t_1,t_2\}$ then
$$\p_AY=\f{\p^2}{\p t_1^2}Y=6t_1t_2^2~~,~~\p_BY=\f{\p^2}{\p t_1\p t_2}Y=6t_1^2t_2$$
For all $0\leq j\leq k$, let $\mb{E}_j(Y)=\max_{|A|\geq
j}\mb{E}(\p_AY)$. Define parameters $c_k,d_k$ recursively as
follows: $~c_1=1,~d_1=2,~c_k=2k^{1/2}(c_{k-1}+1),~d_k=2(d_{k-1}+1)$.

\begin{theorem}[Vu]\label{thm:VuIneq}
Take a polynomial $Y$ as defined above. For any collection of
positive numbers $\mc{E}_0>\mc{E}_1>\cdots>\mc{E}_k=1$ and $\la$
satisfying
\begin{itemize}
\item[(a)] $\mc{E}_j\geq\mb{E}_j(Y)$, and
\item[(b)] $\mc{E}_j/\mc{E}_{j+1}\geq \la+4j\log n,~0\leq j\leq k-1 $
\end{itemize}
the following is true.
$$\mb{P}\left(|Y-\mb{E}(Y)|\geq c_k\sqrt{\la\mc{E}_0\mc{E}_1}\right)\leq d_ke^{-\la/4}.$$
\end{theorem}

%---------------------------------------------------- SUBSUBSECTION PROOFS
\subsubsection{Proof of part (a) of Lemmas \ref{lem:f1A} and \ref{lem:f1Ainf}.}\label{subsubsec:proo-A-lemmas-a}

In order to show part (a) of Lemma \ref{lem:f1A} we
prove the stronger property
\begin{eqnarray}
\mb{P}(A_{i-1}^c)&\leq&
e^{-\Om(\la_i)}.\label{eq:lem-f1A-a-stronger}
\end{eqnarray}
This property combined with $\la_L\geq cd_{\max}^2\log n$ proves part (a)
of Lemma \ref{lem:f1Ainf} as well. From (\ref{eq:T-ineq}) we have
$$A_{i-1}^c\subseteq\{\Psi_r-\psi_r\geq \f{T_r(\la_i)}{8}\}.$$
Hence, in order to show (\ref{eq:lem-f1A-a-stronger}) it is sufficient to show
the following two lemmas.
\begin{lemma}\label{lem:psi>T1}
For all $r$ such that $2m-2r\geq \om\la_i$:
\begin{equation*}
\mb{P}\left(|\Psi_r(\mc{N})-\psi_r|\geq
\f{3\beta_r(\la_i)+2\min(\g_r(\la_i),\nu_r)}{8}\right)\leq e^{-\Om(\la_i)}
\end{equation*}
\end{lemma}
\begin{lemma}\label{lem:psi>T2}
For any $r$ such that $2m-2r<\om\la_i$ we have
$$\mb{P}\left(\Psi_r(\mc{N})-\psi_r\geq\la_i^2/\om\right)\leq e^{-\Om(\la_i)}.$$
\end{lemma}
Now we focus on Lemma \ref{lem:psi>T1}.
For each variable $\Delta_r, \Lambda_r, \Psi_r$ denote their
analogues quantity in $G_{p_r}$ by $\Delta_{p_r}, \Lambda_{p_r},
\Psi_{p_r}$.
\begin{lemma}\label{lem:Gp-numedge}
For all $r$ we have $\mb{P}_{p_r}\big(\{|E(G_{p_r})|=r\}\big)\geq\frac{1}{n}.$
\end{lemma}
\begin{proof}
let $f(m,r)=\mb{P}_{p_r}(\{|E(G_{p_r})|=r\})$ then it can be seen
that
$$\frac{f(m,r+1)}{f(m,r)}=\frac{(1+1/r)^r}{(1+\f{1}{m-r-1})^{m-r}}\leq 1\quad\forall r\leq (m-1)/2.$$
Hence, the minimum of $f(m,r)$ is around $r=m/2$. Using
Stirling's approximation we can get
$f(m,r)\geq\f{1}{\sqrt{2m}}\geq\f{1}{n}$.
\enp\end{proof}
By Lemma
\ref{lem:Gp-numedge}, with probability at least $1/n$, $G_{p_r}$ has
exactly $r$ edges. Hence, using $\la_i\gg\log n$, for proving Lemma
\ref{lem:psi>T1} we only need to show
\begin{equation}\label{eq:Gp-Cons}
\mb{P}\left(|\Psi_{p_r}-\psi_r|\geq\f{3\beta_r(\la_i)+2\min(\g_r(\la_i),~\nu_r)}{8}\right)\leq
e^{-\Om(\la_i)}.
\end{equation}
In order to prove \eqref{eq:Gp-Cons} we define
\begin{eqnarray}
\alpha_r(\la)&=&c\sqrt{\la(m\,d_{\max}q_r^2+\la^2)(d_{\max}q_r+\la)},\no\\
\zeta_r(\la)&=&c\frac{d_{\max}^2}{m}\sqrt{\la(md_{\max}q_r^2+\la^2)(q+\la)}.\no
\end{eqnarray}
It is flashforward that $\alpha_r(\la),~\zeta_r(\la)\leq\beta_r(\la)$.
Therefore, (\ref{eq:Gp-Cons}) is the result of the following lemma.
Throughout the rest of the proof we fix $r,i$ and remove all
sub-indices $r,i$ for simplicity.
%------------------------------------------------------------  MAIN CONCENTRATION LEMMA USING KIM-VU POLYNOMIAL
\begin{lemma}\label{lem:Gp-cons-break} For all $p$ we have
\begin{itemize}
\item[(i)]$\mb{P}\left(|\Delta_p^{(1)}-\mb{E}(\Delta_p^{(1)})|\geq \f{\al}{8}\right)\leq
e^{-\Om(\la)}$
\item[(ii)]$\mb{P}\left(|\Delta_p^{(2)}-\mb{E}(\Delta_p^{(2)})|\geq \f{\min(\beta+\g,\beta+\nu)}{8}\right)\leq
e^{-\Om(\la)}$
\item[(iii)]$\mb{P}\left(|\f{(\Lambda_p^{(1)})^2-\Lambda_p^{(2)}}{8m}-\f{\mb{E}(\Lambda_p^{(1)})^2-\mb{E}(\Lambda_p^{(2)})}{8m}|\geq \f{\zeta}{8}\right)\leq e^{-\Om(\la)}$
\item[(iv)]$\mb{P}\left(|\f{\Lambda_p^{(3)}}{4m}-\f{\mb{E}(\Lambda_p^{(3)})}{4m}|\geq \f{\min(\beta+\g,\beta+\nu)}{8}\right)\leq e^{-\Om(\la)}$
\end{itemize}
\end{lemma}
\begin{proof}
\begin{itemize}
\item[(i)]
Similar to Kim and Vu's proof, for each edge $e$ of $G$ consider a random
variable $t_e$ which is equal to $0$ when $e$ is present in $G_{p}$
and $1$ otherwise. These $t_e$'s will be i.i.d. Bernoulli with mean
$q$. Now note that
$$\Delta_p^{(1)}=\sum_u\sum_{u\in e\cap f,~e\neq f}t_et_f$$
and
$$\mb{E}(\Delta_p^{(1)})=\sum_u{d_u\choose2}q^2\leq m\,d_{\max}q^2.$$
For each $t_e$ we have
$$\mb{E}(\p_{t_e}\Delta_p^{(1)})=\mb{E}(\sum_{f:f\cap
e\neq\emptyset}t_f)\leq2(d_{\max}-1)q<2d_{\max}q.$$
Moreover, any partial
second order derivative is at most 1. Hence,
\begin{eqnarray*}
\mb{E}_0(\Delta_p^{(1)})&\leq&\max(m\,d_{\max}q^2,2d_{\max}q,1),\\
\mb{E}_1(\Delta_p^{(1)})&\leq&\max(2d_{\max}q,1)~\textrm{and},\\
\mb{E}_2(\Delta_p^{(1)})&\leq&1.
\end{eqnarray*}
Now set
$\mc{E}_0=4m\,d_{\max}q^2+4\la^2,~\mc{E}_1=2d_{\max}q+2\la$, and $\mc{E}_2=1$.
Then since $\la\gg\log m$, the conditions of Theorem \ref{thm:VuIneq} are fulfilled.
On the other hand, for $c$ sufficiently large in the definition of $\al$, $c_2\sqrt{\la\mc{E}_0\mc{E}_1}\leq\al/8$.

\item[(ii)] We need to prove the following statements
\begin{eqnarray}
\mb{P}\left(|\Delta_p^{(2)}-\mb{E}(\Delta_p^{(2)})|\geq
\f{\beta+\g}{8}\right)\leq
e^{-\Om(\la)},\label{eq:del2-p-bet-gam}\\
\mb{P}\left(|\Delta_p^{(2)}-\mb{E}(\Delta_p^{(2)})|\geq
\f{\beta+\nu}{8}\right)\leq e^{-\Om(\la)}.\label{eq:del2-p-bet-nu}
\end{eqnarray}
Consider the same random variables $t_e$ from part (i). Let $Q$ be
the set of all paths of length $3$ in $G$. Then
$$\Delta_p^{(2)}=\sum_{\{e,f,g\}\in Q}t_et_g(1-t_f)=\sum_{\{e,f,g\}\in Q}t_et_g-\sum_{\{e,f,g\}\in Q}t_et_ft_g$$
Now let $Y_1 = \sum_{\{e,f,g\}\in Q}t_et_g/4$ and $Y_2=\sum_{\{e,f,g\}\in
Q}t_et_ft_g$. Similar to part (i) we have
$$\mb{E}_0(Y_1)\leq\max(m\,d_{\max}^2q^2/4,d_{\max}^2q/2,1)~,~\mb{E}_1(Y_1)\leq\max(d_{\max}^2q/2,1)~\textrm{and}~\mb{E}_2(Y_1)\leq1.$$
Therefore, set
$\mc{E}_0=m\,d_{\max}^2q^2/2+2\la^2,~\mc{E}_1=d_{\max}^2q/2+2\la$, and $\mc{E}_2=1$.
These satisfy the conditions of Theorem \ref{thm:VuIneq}. Again by
considering $c$ large enough we have
\begin{eqnarray}
\mb{P}(|Y_1-\mb{E}(Y_1)|\geq\beta/32)&\leq&e^{-\Om(\la)}.\label{eq:cons-ii-Y1}
\end{eqnarray}
For $Y_2$ we have
$$\mb{E}_0(Y_2)\leq\max(m\,d_{\max}^2q^3,2d_{\max}^2q^2,2d_{\max}q,1)~,~\mb{E}_1(Y_2)\leq\max(2d_{\max}^2q^2,2d_{\max}q,1)$$
and
$$\mb{E}_2(Y_2)\leq\max(2d_{\max}q,1)~\textrm{and}~\mb{E}_3(Y_2)=1.$$
As before, set
$\mc{E}_0=2m\,d_{\max}^2q^3+3\la^3$, $\mc{E}_1=2d_{\max}^2q^2+2\la^2$, and $\mc{E}_2=2d_{\max}q+\la,\mc{E}_3=1$ to obtain
\begin{eqnarray}
\mb{P}(|Y_2-\mb{E}(Y_2)|\geq\f{\g}{8})&\leq&e^{-\Om(\la)}.\label{eq:cons-ii-Y2}
\end{eqnarray}
Combining (\ref{eq:cons-ii-Y1}) and (\ref{eq:cons-ii-Y2}), equation (\ref{eq:del2-p-bet-gam}) is proved.
Finally, equation (\ref{eq:del2-p-bet-nu}) is the result of (\ref{eq:cons-ii-Y1})
and the following,
\begin{eqnarray*}
|\Delta_p^{(2)}-\mb{E}(\Delta_p^{(2)})|&\leq& |4Y_1-4\mb{E}(Y_1)|+\mb{E}(Y_2)\\
&\leq& |4Y_1-4\mb{E}(Y_1)|+m\,d_{\max}^2q^3\\
&\leq& |4Y_1-4\mb{E}(Y_1)|+\f{\nu}{8}.
\end{eqnarray*}

\item[(iii)] Here we will prove
\begin{eqnarray}
\mb{P}\left(|\f{(\Lambda_p^{(1)})^2}{8m}-\f{\mb{E}(\Lambda_p^{(1)})^2}{8m}|\geq
c_1d_{\max}^2q\sqrt{\la(\la+mq)}\right)\leq e^{-\Om(\la)},\label{eq:cons-iii-1}
\end{eqnarray}
and
\begin{eqnarray}
\mb{P}\left(|\f{\Lambda_p^{(2)}}{8m}-\mb{E}(\f{\Lambda_p^{(2)}}{8m})|\geq\f{c_1d_{\max}^2}{m}\sqrt{\la(m d_{\max}q^2+2\la^2)(q+\la)}\right)\leq
e^{-\Om(\la)}.\label{eq:cons-iii-2}
\end{eqnarray}
Note that by making $c$ in the definition of $\zeta$ large enough,
(\ref{eq:cons-iii-1}) and (\ref{eq:cons-iii-2}) together give us
(iii). First we prove (\ref{eq:cons-iii-1}). Write
$$\f{\Lambda_p^{(1)}}{2d_{\max}}=\sum_{e=\{u,v\}\in E(G)}\f{d_u+d_v}{2d_{\max}}t_e$$
which is a polynomial with coefficients in $(0,1]$. As before
$$\mb{E}_0(\f{\Lambda_p^{(1)}}{2d_{\max}})\leq\max(mq,1)~~,~~\mb{E}_1(\f{\Lambda_p^{(1)}}{2d_{\max}})\leq1.$$
Now set $\mc{E}_0=\la+mq$ and $\mc{E}_1=1$. Thus,
\begin{eqnarray}
\mb{P}\left(|\f{\Lambda_p^{(1)}}{2d_{\max}}-\mb{E}(\f{\Lambda_p^{(1)}}{2d_{\max}})|\leq
c_1\sqrt{\la(\la+mq)}\right)\leq d_1e^{-\Om(\la)}.\no\\
\label{eq:cons-iii-L1}
\end{eqnarray}
By Lemma \ref{lem:del-lam-bounds}(ii) we have $\Lambda_p^{(1)}\leq
2m\,d_{\max}q$. Hence, inequality $|(\Lambda_p^{(1)})^2-\mb{E}(\Lambda_p^{(1)})^2|\geq
8c_1m\,d_{\max}^2q\sqrt{\la(\la+mq)}$ gives
$$|\f{\Lambda_p^{(1)}}{2d_{\max}}-\mb{E}(\f{\Lambda_p^{(1)}}{2d_{\max}})|\geq
c_1\sqrt{\la(\la+mq)}.$$
Now using (\ref{eq:cons-iii-L1}) equation (\ref{eq:cons-iii-1}) is trivial.

The proof of (\ref{eq:cons-iii-2}) is similar to the proofs in (i) and (ii). We start with the following polynomial representation for $\Lambda_p^{(2)}$
\begin{eqnarray}
\f{\Lambda_p^{(2)}}{2d_{\max}^2}&=&\sum_{i=1}^n\frac{d_i^2}{2d_{\max}^2}\left(\sum_{e=(i,.)}t_e\right)^2\no\\
&=&\sum_{i=1}^n \frac{d_i^2}{2d_{\max}^2}\left(\sum_{e=(i,.)}t_e\right)+2\sum_{i=1}^n \frac{d_i^2}{2d_{\max}^2}\sum_{e\cap f=i}t_et_f\no.
\end{eqnarray}
Then we represent the right hand side by $Z_1+Z_2$ where $Z_1=\sum_{i=1}^n \frac{d_i^2}{2d_{\max}^2}\left(\sum_{e=(i,.)}t_e\right)$
and $Z_2=2\sum_{i=1}^n \frac{d_i^2}{2d_{\max}^2}\sum_{e\cap f=i}t_et_f$. The next step is to use Vu's inequality for both $Z_1$ and $Z_2$ separately. The concentration for $Z_2$ is less sharp and it will dominate the concentration for $Z_1+Z_2$.
For $Z_1$ the inequalities
$$\mb{E}_0(Z_1)\leq \max(mq,1)~~,~~\mb{E}_1(Z_2)\leq 1$$
show that the same $\mc{E}_0,\mc{E}_1$ as in \eqref{eq:cons-iii-L1} can be used to obtain the inequality
\begin{eqnarray}
\mb{P}\left(|\frac{2d_{\max}^2Z_1}{8m}-\mb{E}(\frac{2d_{\max}^2Z_1}{8m})|\leq c_2\frac{d_{\max}^2}{m}\sqrt{\la(\la+mq)}\right)
\leq d_2e^{-\Om(\la)}.\no\\
\label{eq:cons-iv-Z1}
\end{eqnarray}
Now for $Z_2$ the bounds on the partial derivatives are given by
$\mb{E}_0(Z_2)\leq\max(\frac{m\,d_{\max}q^2}{2},q,1)$, $\mb{E}_1(Y_1)\leq\max(q,1)$, and $\mb{E}_2(Y_1)=1.$
Therefore, $\mc{E}_0=m\,d_{\max}q^2+2\la^2$ and $\mc{E}_1=q+\la,\mc{E}_2=1$ satisfy the conditions of Theorem \ref{thm:VuIneq} and we obtain the inequality
\begin{eqnarray}
\mb{P}\left(|\frac{2d_{\max}^2Z_2}{8m}-\mb{E}(\frac{2d_{\max}^2Z_2}{8m})|\leq c_3\frac{d_{\max}^2}{m}\sqrt{\la(m\,d_{\max}q^2+2\la^2)(q+\la)}\right)
\leq d_2e^{-\Om(\la)}.\no\\
\label{eq:cons-iv-Z2}
\end{eqnarray}
The final inequality \eqref{eq:cons-iii-2} can now be shown by combining equations \eqref{eq:cons-iv-Z1} and \eqref{eq:cons-iv-Z2}.

\item[(iv)] This case is treated exactly the same as (ii) because we have the following
$$\f{\Lambda_p^{(3)}}{d_{\max}^2}=\sum_{\{e,f,g\}\in R,~e=\{u,v\}}\f{d_ud_v}{d_{\max}^2}t_et_g(1-t_f).$$
\end{itemize}

\enp\end{proof}
%----------------------------------------  PROOF OF LEMMA Psi diffs > T2
\begin{proof}[of Lemma \ref{lem:psi>T2}]
Using Lemma \ref{lem:del-lam-bounds}(iii) and the definition of $\Psi$,
from $\Psi_p\geq\la^2/\om$ we can get
\begin{eqnarray}
\Delta_p&\geq&\f{\la^2}{\om}-\Lambda_p\no\\%\f{(\Lambda_p^{(1)})^2+\Lambda_p^{(2)}}{8m}+\f{\Lambda_p^{(3)}}{4m}\no\\
&\geq&\f{\la^2}{\om}-m\,d_{\max}^2q^2\no\\
&>&\f{\la^2}{\om}- \f{d_{\max}^2\om^2\la^2}{4m}\label{eq:use-q<omla}\\
&>&\f{\la^2}{2\om}\label{eq:use-d4=om}
\end{eqnarray}
where (\ref{eq:use-q<omla}) uses $2mq=2m-2r<\om\la$ and
(\ref{eq:use-d4=om}) holds since $d_{\max}^2\om^3\ll m$.

Since $2m-2r$ is small then $G_p$ is very dense. Let us consider its
complement $G_q$ which is sparse. Let $N_0(u)=N(u)\cup\{u\}$. Then
using
$$\Delta_{p_r}\leq\sum_u d_{G_q}(u)\sum_{v\in N_0(u)}d_{G_q}(v)$$
and $\Delta_p\geq\la^2/2\om$, one of the following statements should hold.
\begin{itemize}
\item[(a)] $G_q$ has more than $\om^2\la/4$ edges.
\item[(b)] For some $u$, $\sum_{v\in N_0(u)}d_{G_q}(v)\geq\la/\om^3$.
\end{itemize}
If (a) holds, since $2mq\leq \om\la$ then
\begin{eqnarray}
\mb{P}(G_q\textrm{ has more than $\f{\om^2\la}{4}$
edges})&\leq&{m\choose\f{\om^2\la}{4}}q^{\f{\om^2\la}{4}}\no\\
&\leq&(\f{4mqe}{\om^2\la
})^{\f{\om^2\la}{4}}\no\\
&\leq &e^{-\f{\om^2\la}{4}(\log\om-1-\log2)}
=e^{-\Om(\la)}.\no
\end{eqnarray}
If (b) holds then the number of edges in $G$ that contribute to
$\sum_{v\in N_0(u)}d_{G_q}(v)$ is at most $d_{\max}^2$ and each edge can
contribute at most twice. Hence,
\begin{eqnarray}
\mb{P}(\sum_{v\in
N_0(u)}d_{G_q}(v)\geq\la/\om^3)&\leq&{d_{\max}^2\choose\f{\la}{2\om^3}}q^{\f{\la}{2\om^3}}\no\\
&\leq&
(\f{2d_{\max}^2q\om^3e}{\la})^{\f{\la}{2\om^3}}\no\\
&\leq&
(\f{d_{\max}^2\om^4e}{m})^{\f{\la}{2\om^3}}\no\\
&=& e^{-\f{\la}{2\om^3}(\log m-\log(d_{\max}^2\om^4)-1)}\leq
e^{-\Om(\f{\la}{\om^3}\log m)}=e^{-\Om(\la)}.\no
\end{eqnarray}
Note that we need $\delta$ in the definition of $\om$ to be small enough
such that $\log m\gg\om^3$ and for $\delta < .1$ this is
true.
\enp\end{proof}

\subsubsection{Proof of part (b) of Lemmas \ref{lem:f1A} and
\ref{lem:f1Ainf}.}\label{subsubsec:proo-A-lemmas-b}

Note that:
$$f(\mc{N})=\prod_{r=0}^{m-1}\left(1+\f{\Psi_r(\mc{N})-\psi_r}
{{2m-2r\choose2}-\Psi_r(\mc{N})}\right)
$$
and since $\Psi_r(\mc{N})\leq(1-\tau/4){2m-2r\choose2}$ for
$\mc{N}\in S^*(\mc{M})$ then
$$f(\mc{N})\leq\prod_{r=0}^{m-1}\left(1+\f{16/\tau\max(\Psi_r(\mc{N})-\psi_r,0)}
{(2m-2r)^2}\right).
$$
\begin{proof}[of Lemma \ref{lem:f1A}(b)]
Using $1+x\leq e^x$ we only need to show
$$\sum_{r=0}^{m-1}\f{\max(\Psi_r(\mc{N})-\psi_r,0)}
{(2m-2r)^2}\leq o(\la).
$$
To simplify the notation, let $g(r)=
\f{\max(\Psi_r(\mc{N})-\psi_r,0)}{(2m-2r)^2}$. Note that $0\leq
g(r)\leq 1$ which gives $\sum_{2m-2r=2}^{\la/\om^{1/2}}g(r)=o(\la)$. Hence, we
only need to show $\sum_{2m-2r=\la/\om^{1/2}}^{2m-2}g(r)=o(\la)$ .
Also note that the numerator of $g(r)$ is at most $T_r(\la)$. Therefore, using the
definition of $T_r(\la)$,
\begin{eqnarray*}
\sum_{2m-2r=\la/\om^{1/2}}^{2m-2}g(r)&\leq&\sum_{2m-2r=\la/\om^{1/2}}^{\om\la}\f{\la^2}{(2m-2r)^2\om}+\sum_{2m-2r=\om\la}^{\om\la^2}\frac{3\beta_r(\la)+2\nu_r}{(2m-2r)^2}\no\\
&&~~~~~~~~~~+\sum_{2m-2r=\om\la^2}^{2m-2}\frac{3\beta_r(\la)+2\g_r(\la)}{(2m-2r)^2}.
\end{eqnarray*}
Therefore, it suffices to show
\begin{multline*}
\sum_{2m-2r=\la/\om^{1/2}}^{\om\la}\f{\la^2}{(2m-2r)^2\om}
+\sum_{2m-2r=\om\la}^{\om\la^2}\frac{3\beta_r(\la)+2\nu_r}{(2m-2r)^2}\\
+\sum_{2m-2r=\om\la^2}^{2m-2}\frac{3\beta_r(\la)+2\g_r(\la)}{(2m-2r)^2}=o(\lambda)
\end{multline*}
A series of elementary inequalities will now be used to bound these
three summations. We will use $q_r=\f{2m-2r}{2m}$ to obtain
\begin{eqnarray*}
\sum_{2m-2r=2}^{2m-2}\f{(\la
m\,d_{\max}^4q_r^3)^{1/2}}{(2m-2r)^2}&=&\f{\la^{1/2}d_{\max}^2}{2m\sqrt{2}}\sum_{2m-2r=2}^{2m-2}\f{1}{\sqrt{2m-2r}}\\
&=&O(\f{\la^{1/2}d_{\max}^2}{m}\int_{x=2}^{2m}\f{1}{\sqrt{x}}dx)=O(\f{\la^{1/2}d_{\max}^2}{\sqrt{m}})=o(\la)
\end{eqnarray*}
$$
\sum_{2m-2r=2}^{2m-2}\f{(\la^2
m\,d_{\max}^2q_r^2)^{1/2}}{(2m-2r)^2}=\f{\la
d_{\max}}{2m^{1/2}}\sum_{2m-2r=2}^{2m-2}\f{1}{2m-2r}=O(\f{\la
d_{\max}}{m^{1/2}}\log m)=o(\la),
$$
$$
\sum_{2m-2r=2}^{2m-2}\f{(\la^3
d_{\max}^2q_r)^{1/2}}{(2m-2r)^2}=\f{\la^{3/2}
d_{\max}}{(2m)^{1/2}}\sum_{2m-2r=2}^{2m-2}\f{1}{(2m-2r)^{3/2}}=O(\f{\la^{3/2}
d_{\max}}{m^{1/2}})=o(\la),
$$
and
$$
\sum_{2m-2r=\om\la}^{2m-2}\f{\la^2}{(2m-2r)^2}\leq\la^2\int_{x=\om\la}^{2m}x^{-2}dx=o(\la).
$$
Furthermore, we can show the following bounds
$$
\sum_{2m-2r=2}^{2m-2}\f{(\la^3
m\,d_{\max}^2q_r^3)^{1/2}}{(2m-2r)^2}=
\f{\la^{3/2}d_{\max}}{2m\sqrt{2}}\sum_{2m-2r=2}^{2m-2}\f{1}{\sqrt{2m-2r}}=O(\f{\la^{3/2}d_{\max}}{\sqrt{m}}),
$$
$$
\sum_{2m-2r=2}^{2m-2}\f{\la^2d_{\max}q_r}{(2m-2r)^2}=O(\f{\la^2d_{\max}\log
m}{2m}),
$$
\begin{eqnarray}
\sum_{2m-2r=\om\la^2}^{2m-2}\f{\la^3}{(2m-2r)^2}=O(\la^3\int_{x=\om\la^2}^{\infty}x^{-2}dx)=O(\f{\la^3}{\om\la^2})=o(\la)\label{eq:29},
\end{eqnarray}
and
\begin{eqnarray}
\sum_{2m-2r=2}^{\om\la^2}\f{m\,d_{\max}^2q_r^3}{(2m-2r)^2}=
\sum_{2m-2r=2}^{\om\la^2}\f{d_{\max}^2(2m-2r)}{8m^2}=O(\f{\om^2\la^4d_{\max}^2}{m^2}).\label{eq:30}
\end{eqnarray}
\begin{remark} All previous equations are of order $o(\la)$, since
$\la\leq\la_L=O(d_{\max}^2\log n)$ and
$d_{\max}=o(m^{\f{1}{4}-\tau})$. Note that we also used $\sqrt{A+B}\leq\sqrt{A}+\sqrt{B}$ to find upper bounds for
$\beta_r,\g_r$.
\end{remark}
\enp\end{proof}

\begin{proof}[of Lemma \ref{lem:f1Ainf}(b)]
Similar to proof of Lemma \ref{lem:f1A}(b) we will show
\begin{eqnarray}
f(\mc{N})&\leq&\prod_{r=m-d_{\max}^2+1}^m\f{{2m-2r\choose
2}-\psi_r}{{2m-2r\choose
2}-\Psi_r(\mc{N})}\prod_{r=0}^{m-d_{\max}^2}\big(1+16/\tau\f{\max(\Psi_r(\mc{N})-\psi_r,0)}
{(2m-2r)^2}\big)\no\\
&\leq&{2d_{\max}^2\choose
2}^{d_{\max}^2}\cdot\prod_{r=0}^{m-d_{\max}^2}\big(1+16/\tau\f{\Psi_r}{(2m-2r)^2}\big)\no\\
&\leq&(2d_{\max}^4)^{d_{\max}^2}\cdot\prod_{r=0}^{m-d_{\max}^2}\big(1+16/\tau\f{d_{\max}^2}{2m-2r}\big)\label{eq:twenty}\\
&\leq&e^{d_{\max}^2\log(2d_{\max}^4) +
3\sum_{i=d_{\max}^2+1}^m\f{d_{\max}^2}{i}}\no\\
&\leq&e^{d_{\max}^2\big(\log(2d_{\max}^4)+3\log d_{\max} +
\log m\big)}\no\\
&\leq&e^{4d_{\max}^2\log n}\label{eq:twenty1}
\end{eqnarray}
where (\ref{eq:twenty}) use Lemma \ref{lem:del-lam-bounds}, and
(\ref{eq:twenty1}) uses $m\leq nd_{\max}/2$ and $d_{\max}\ll m^{1/3}\leq
n^{1/2}$.
\enp\end{proof}

\subsubsection{Proof of Lemma \ref{lem:f1C}.}\label{subsubsec:lem-C-proof}

By the definition of $\mc{C}:~\sum_{2m-2r=2}^{\om\la_0}g(r)=0$. Thus, we only
need to show that if $\Psi_r(\mc{N})-\psi_r\leq T_r(\la_0)$ for
all $r$ with $2m-2r\geq\om\la_0$ then
$$\sum_{2m-2r=\om\la_0}^{m}g(r)=o(1).$$
For that it is sufficient to prove
$$
\sum_{2m-2r=\om\la_0}^{m}\f{T_r(\la_0)}{(2m-2r)^2}=o(1).
$$
The proof is similar to the proof of Lemma \ref{lem:f1A} (b) with a
slight modification. Instead of using (\ref{eq:29}) and (\ref{eq:30}) we use
$$
\sum_{2m-2r=\om\la_0^3}^{2m-2}\f{\la_0^3}{(2m-2r)^2}=O(\la_0^3\int_{\om\la_0^3}^\infty
x^{-2}dx)=O(\f{\la_0^3}{\om\la_0^3})=o(1),
$$
and
$$
\sum_{2m-2r=2}^{\om\la_0^3}\f{m\,d_{\max}^2q_r^3}{(2m-2r)^2}=
\sum_{2m-2r=2}^{\om\la_0^3}\f{(2m-2r)d_{\max}^2}{m^2}=O(\f{d_{\max}^2\om^2\la_0^6}{m^2})=o(1).
$$
For the other equations in the proof of Lemma \ref{lem:f1A} (b) let $\la=\la_0$ and they will be $o(1)$.

\subsubsection{Proof of Lemma \ref{lem:f1B}.}\label{subsubsec:lem-B-proof}

\begin{proof}[of Lemma \ref{lem:f1B}(a)]
We have $2m-2r\leq\om\la_0\ll (\log n)^2$. This means proving the
bound only for one $r$ is enough. Similar to the proof of Lemma
\ref{lem:psi>T2}, from $\Psi_p\geq 2^{j-1}$ we get $\Delta_p\geq
2^{j-2}$. Thus, one of the following statements hold
\begin{itemize}
\item[(a)] $G_q$ has more than $2^{j/2-2}$ edges
\item[(b)] For some $u$, $\sum_{v\in N_0(u)}d_{G_q}(v)\geq2^{j/2-1}$
\end{itemize}
and rest of the proof will be exactly as in Lemma \ref{lem:psi>T2}.
\enp\end{proof}

\begin{proof}[of Lemma \ref{lem:f1B}(b)]
By the definition of $B_j$
$$
\sum_{2m-2r=2}^{\om\la_0}g(r)\leq
\sum_{2m-2r=2}^{\om\la_0}\f{2^j}{(2m-2r)^2}=O(2^j).
$$
\enp\end{proof}

%-------------------------------------------------  SUBSECTION EASY LOWER BOUND
\subsection{Proof of (\ref{eq:av-f1C>1-o1})}\label{subsec:pf-av-f1C>1-o1}

From Lemma \ref{lem:psi>T1}, for all $r$ with $2m-2r\geq\om\la_0$,
\begin{eqnarray}
\mb{P}\Big(|\Psi_r-\psi_r|\geq\al_r(\la_0)+\beta_r(\la_0)+(1+d_{\max}^2/4m)\g_r(\la_0)+\zeta_r(\la_0)\Big)=
o(1).\label{eq:Psi-psi>...=o1}
\end{eqnarray}
Let $\mc{N}$ be an ordering with
$|\Psi_r-\psi_r|\leq\al_r(\la_0)+\beta_r(\la_0)+(1+d_{\max}^2/4m)\g_r(\la_0)+\zeta_r(\la_0)$
for all $2m-2r\geq\om\la_0$. Then
\begin{eqnarray}
f(\mc{N})&\geq&\prod_{2m-2r=\om\la_0^3}^{2m-2}\left(1-(16/\tau)\f{\al_r(\la_0)+\beta_r(\la_0)+\g_r(\la_0)+\zeta_r(\la_0)}{(2m-2r)^2}\right)\no\\
&&\times\prod_{2m-2r=2}^{\om\la_0^3}\left(1-(16/\tau)\f{\psi_r}{(2m-2r)^2}\right).\label{eq:f(N)>=Prod1xProd2}
\end{eqnarray}
In section \ref{subsubsec:proo-A-lemmas-b} it was shown that
$\f{3}{\tau}\sum_{2m-2r=\om\la_0^3}^{2m-2}\f{\al_r(\la_0)+\beta_r(\la_0)+\g_r(\la_0)+\zeta_r(\la_0)}{(2m-2r)^2}=o(1)$.
Now one can use $1-x\geq e^{-2x}$ when $0\leq x\leq1/2$ to see that the
first product in the right hand side of \eqref{eq:f(N)>=Prod1xProd2} is $1-o(1)$. The second product is also
$1-o(1)$ because of $\om\la_0^3d^2=o(m)$ and the bound $\psi_r=O\left[(2m-2r)^2\f{d_{\max}^2}{m}\right]$ given Lemma \ref{lem:bound-on-psi}. These, together with (\ref{eq:Psi-psi>...=o1})
finish the proof of (\ref{eq:av-f1C>1-o1}). In fact they show the
stronger statement $\mb{E}(f(\mc{N})\mrm{1}_{S^*(\mc{M})})>1-o(1)$.

\begin{remark} The proofs of this section and Section \ref{subsec:four-lem} yield the
following corollary which will be used in Section
\ref{subsec:pf-S*c-o1}.
\end{remark}

\begin{corollary}\label{cor:av-ep-gr} For sufficiently large $c$ in the definition of $\la_L$,
\begin{eqnarray}
\mb{E}\left(\exp\bigg[{\f{1}{\tau^2}\sum_{r=0}^{m-1}\f{\max(\Psi_r(\mc{N})-\psi_r,0)}{(2m-2r)^2}}\bigg]\right)&=&1+o(1)
\end{eqnarray}
\end{corollary}
\begin{proof}
Bounds of Section \ref{subsec:four-lem} show
that the contribution of the sets $A_i\stm A_{i-1}$ and $B_j\stm
B_{j-1}$ are all $o(1)$ and the contribution of $\mc{C}$ is $1+o(1)$.
The contribution of $A_\infty$ also is $o(1)$ by taking the constant $c$
large enough.
\enp
\end{proof}
%---------------------------------------------------------------- SUBSECTION PROOF OF o(1)
\subsection{Proof of
(\ref{eq:av-f1S*c=o1})}\label{subsec:pf-S*c-o1}

In this section we deal with those orderings $\mc{N}$ for which the
condition
$$\Psi_r(\mc{N})\leq(1-\tau/4){2m-2r\choose2}\quad\quad\quad(\ast)$$
is violated for some $r$. If this happens for some $r$ then from
Lemma \ref{lem:del-lam-bounds}(iii) and $d_{\max}^4=o(m)$ we have
\begin{multline*}
\Delta_r(\mc{N})\geq\Psi_r(\mc{N})-\f{d_{\max}^2}{8m}(2m-2r)^2\\
>\Psi_r(\mc{N})-\tau/4{2m-2r\choose2}>(1-\tau/2){2m-2r\choose2}.
\end{multline*}
On the other hand using Lemma \ref{lem:del-lam-bounds}(i) we have
$\Delta_r(\mc{N})\leq \f{d_{\max}^2(2m-2r)}{2}$.  So for
$2m-2r\geq\f{d_{\max}^2}{2-\tau}$ we have
$\Delta_r(\mc{N})\leq(1-\tau/2){2m-2r\choose2}$. Thus condition
$(*)$ is violated only for $r$ very close to $m$. Let
$S_t(\mc{M}),~t=1,\ldots,\f{d_{\max}^2}{2-\tau}$, be the set of
all ordering $\mc{N}$ for which $(*)$ fails for the first time at
$r=m-t$. We will use $\sum_{t=1}^\infty\f{1}{m^{\tau t}}=o(1)$ to
prove (\ref{eq:av-f1S*c=o1}). In particular we show
$$\mb{E}(f(\mc{N})\mrm{1}_{S_t})\leq O\Big(\f{1}{m^{\tau t}}\Big).$$
Note that ${2m-2r\choose2}-\Psi_r(\mc{N})=
\sum_{\{i,j\}~\in~E_r}d_i^{(r)}d_j^{(r)}(1-\f{d_id_j}{4m})\geq
(m-r)(1-\f{d_{\max}^2}{4m})$ since at step $r$ there should be at
least $m-r$ suitable edges to complete the ordering $\mc{N}$. Hence
using
$d_{\max}=O(m^{\f{1}{4}-\tau})$ we have%------------------------------------------------------- CONSTRAINT d=o(m^{\f{1}{4}-\tau}) IS USED
\begin{equation}\label{eq:case(m-r)<d}
\f{{2m-2r\choose2}}{{2m-2r\choose2}-\Psi_r(\mc{N})}\leq
2m-2r-1+O(\f{d_{\max}^4}{m})\leq 2m-2r.
\end{equation}
This gives
$$
\prod_{r=m-t}^{m-1}\f{{2m-2r\choose2}}{{2m-2r\choose2}-\Psi_r(\mc{N})}\leq
2^tt!\leq2t^t
$$
and since $t$ is the first place that $(*)$ is violated, then
$$
\prod_{r=0}^{m-t-1}\f{{2m-2r\choose2}-\psi_r}{{2m-2r\choose2}-\Psi_r(\mc{N})}\leq
\exp\bigg[{\f{16}{\tau}\sum_{r=0}^{m-1}\f{\max(\Psi_r(\mc{N})-\psi_r,0)}{(2m-2r)^2}}\bigg].
$$
Thus,
$$
f(\mc{N})\mrm{1}_{S_t}=\mrm{1}_{S_t}\prod_{r=0}^{m-1}\f{{2m-2r\choose2}-\psi_r}{{2m-2r\choose2}-\Psi_r(\mc{N})}\leq
2t^t\mrm{1}_{S_t}\exp\bigg[{\f{16}{\tau}\sum_{r=0}^{m-1}\f{\max(\Psi_r(\mc{N})-\psi_r,0)}{(2m-2r)^2}}\bigg].
$$
Now using H\"{o}lder's inequality
\begin{eqnarray}
\mb{E}(f(\mc{N})\mrm{1}_{S_t})&\leq&2t^t\mb{E}\left(\mrm{1}_{S_t}\exp\bigg[{\f{16}{\tau}\sum_{r=0}^{m-t-1}\f{\max(\Psi_r(\mc{N})-\psi_r,0)}{(2m-2r)^2}}\bigg]\right)\no\\
&\leq&2t^t\mb{E}(\mrm{1}_{S_t})^{1-\tau/2}\mb{E}\left(\mrm{1}_{S_t}\exp\bigg[{\f{32}{\tau^2}\sum_{r=0}^{m-t-1}\f{\max(\Psi_r(\mc{N})-\psi_r,0)}{(2m-2r)^2}}\bigg]\right)^{\tau/2}.\no
\end{eqnarray}
But using Corollary \ref{cor:av-ep-gr}, the second term
in the above product is $1+o(1)$ and we only need to show
$$
2t^t\mb{P}(S_t)^{1-\tau/2}\leq\Big(1+o(1)\Big)\f{1}{m^{\tau t}}.
$$
Let $r=m-t$ and $\G(u)=N_{G_{\mc{N}_r}}(u)$ be the set of all neighbors
of $u$ in $G_{\mc{N}_r}$. Note that
$$
\Delta_r(\mc{N})=\f{1}{2}\sum_ud_u^{(r)}\sum_{v\in\G(u)\cup\{u\}}(d_v^{(r)}-1_{u=v})
$$
and
$$
{2m-2r\choose2}=\f{1}{2}\sum_ud_u^{(r)}\sum_v(d_v^{(r)}-1_{u=v}).
$$
Now
$\Delta_r(\mc{N})>(1-\tau/2){2m-2r\choose2}>(1-\tau){2m-2r\choose2}
$ implies that a vertex $u$ with $d_u^{(r)}>0$ exists and
$$
\sum_{v\in\G(u)\cup\{u\}}(d_v^{(r)}-1_{u=v})>(1-\tau)\sum_v(d_v^{(r)}-1_{u=v}).
$$
Equivalently
\begin{eqnarray}
\sum_{v\notin\G(u)\cup\{u\}}d_v^{(r)}\leq\tau\sum_v(d_v^{(r)}-1_{u=v})
\leq\tau(2m-2r-1)\leq2\tau t. \label{eq:Verylast}
\end{eqnarray}
Any of the last $t$ edges of $\mc{N}$ that that have at least one endpoint outside of $\G(u)$,
contributes at least once to the left hand side of (\ref{eq:Verylast}). So there are at most $2\tau t$ such
edges. Let $k=d_u-|\G(u)|$ and let $\ell$ be the number of edges that are entirely in
$\G(u)$. Then we should have $k\geq 1$ and $\ell\geq(1-2\tau)i$. Thus, the
probability that $d_u^{(r)}>0$ and
$\sum_{v\notin\G(u)\cup\{u\}}d_v^{(r)}\leq2\tau t$, for a fixed vertex
$u$ is upper bounded by
$$\sum_{k\geq1,~\ell\geq(1-2\tau)t}\f{{d_u\choose k}{{d_u-k\choose2}\choose\ell}{m-d_u-{d_u-k\choose2}\choose t-k-\ell}}{{m
\choose t}}.
$$
Hence,
$$
\mb{P}(S_t)\leq\sum_u\sum_{k\geq1,~\ell\geq(1-2\tau)t}\f{{d_u\choose
k}{{d_u-k\choose2}\choose\ell}{m-d_u-{d_u-k\choose2}\choose
t-k-\ell}}{{m \choose t}}.
$$
Now using
$${d_u\choose k}\leq\f{d_u^k}{k!},~~{{d_u-k\choose2}\choose\ell}\leq\f{(d_u^2/2)^\ell}{\ell!},$$
$${m-d_u-{d_u-k\choose2}\choose t-k-\ell}\leq\f{m^{t-k-\ell}}{(t-k-\ell)!}$$
for $t=O(d_{\max}^2)=o(m^{1/2})$ we have
$${m\choose t}=\big(1+o(1)\big)\f{m^t}{t!}.$$
This means
\begin{eqnarray}
\mb{P}(S_t)&\leq&\big(1+o(1)\big)\sum_u\sum_{k\geq1,~\ell\geq(1-2\tau)t}\f{\f{d_u^k}{k!}\f{(d_u^2/2)^\ell}{\ell!}\f{m^{t-k-\ell}}{(t-k-\ell)!}}{\f{m^t}{t!}}\no\\
&=&\big(1+o(1)\big)\sum_u\sum_{k\geq1,~\ell\geq(1-2\tau)t}\f{(d_u/m)^k(d_u^2/2m)^\ell t!}{k!\ell!(t-k-\ell)!}\no\\
&\leq&\big(1+o(1)\big)2\tau t\sum_u(d_u/m)(d_u^2/2m)^{(1-2\tau)t}{t\choose2\tau t}\no\\
&\leq&\big(1+o(1)\big)t\f{d_{\max}}{m}\sum_u(d_u^2/2m)^{(1-2\tau)t}2^t\label{eq:us-ep-bd-1}\\
&\leq&\big(1+o(1)\big)t2^{2t/3}\f{d_{\max}}{m}\sum_u\left(\f{d_u^2}{m}\right)^{(1-2\tau)t}\label{eq:us-ep-bd-2}\\
&\leq&\big(1+o(1)\big)2t2^{2t/3}\left(\f{d_{\max}^2}{m}\right)^{(1-2\tau)t}\label{eq:sum-d/m}
\end{eqnarray}
where (\ref{eq:us-ep-bd-1}) and (\ref{eq:us-ep-bd-2}) are based on $\tau\leq1/3$
and ${a\choose b}\leq 2^a$.  Moreover, (\ref{eq:sum-d/m}) uses $\sum_u
d_u^k=\sum_{u\sim_G v}(d_u^{k-1}+d_v^{k-1})\leq 2m\,d_{\max}^{k-1}$.
Now we can use $t\leq\f{d_{\max}^2}{2-\tau}$, $d_{\max}\leq
m^{\f{1}{4}-\tau}$, and $\tau\leq 1/3$ to get
\begin{eqnarray}
2t^t\mb{P}(S_t)^{1-\tau/2}&\leq&\big(1+o(1)\big)4t\left(\f{2^{2-\tau/3}}{2-\tau}\f{d_{\max}^{4-5\tau+2\tau^2}}{m^{1-2.5\tau+\tau^2}}\right)^t\no\\
&\leq&\big(1+o(1)\big)4t\left(\f{d_{\max}^{4-5\tau+2\tau^2}}{m^{1-2.5\tau+\tau^2}}\right)^t\no\\
&\leq&\big(1+o(1)\big)4t\left(m^{-2.75\tau+3.5\tau^2-2\tau^3}\right)^t\no\\
&\leq&O\big(m^{-\tau t}\big).\no
\end{eqnarray}
\enp

%----------------------------------------------------------  Bounding the variances.
\section{Bounding the Variance of the SIS estimate}\label{sec:rg-bound-variance}
In this section we will prove two variance bounds from Section \ref{sec:analysis}.
We will borrow some notation and results from Section \ref{sec:Kim-Vu-Calc}.
\subsection{Proof of Equation (\ref{eq:varN=o(1)}) }\label{subsec:varN=o(1)}
It is easy to see that instead of proving (\ref{eq:varN=o(1)}) directly, we can consider the equivalent formulation $\mb{E}_{\mrm{A}}(N^2)/\mb{E}_{\mrm{A}}(N)^2\leq 1+o(1)$.
For the numerator we have
$$\mb{E}_\mrm{A}(N^2)=\sum_G\sum_{\mc{N}}\left(\f{1}{m!~\mb{P}_{\mrm{A}}(\mc{N})}\right)^2\mb{P}_{\mrm{A}}(\mc{N})=\sum_G\sum_{\mc{N}}\f{1}{(m!)^2~\mb{P}_{\mrm{A}}(\mc{N})}.$$
On the other, we have the following estimate from the analysis of Theorem \ref{thm:main-nobias},
$$|\mc{L}(\bar{d})|=\frac{[1+o(1)]\prod_{r=0}^{m-1}\left[{2m-2r\choose2}-\psi_r\right]}{m!\prod_{i=1}^nd_i!}.$$
Therefore,
\begin{eqnarray}
\f{\mb{E}_{\mrm{A}}(N^2)}{\mb{E}_{\mrm{A}}(N)^2}&=&\f{\sum_G\sum_{\mc{N}}\f{1}{(m!)^2~\mb{P}_{\mrm{A}}(\mc{N})}}{|\mc{L}(\bar{d})|^2}\no\\
&=&\f{\sum_G\sum_{\mc{N}}\prod_{r=0}^{m-1}\left[\frac{{2m-2r\choose2}-\Psi_r(\mc{N})}{{2m-2r\choose2}-\psi_r}\right]}{m!|\mc{L}(\bar{d})|}\no\\
&=&\f{\sum_G\mb{E}(g(\mc{N}))}{|\mc{L}(\bar{d})|}\label{eq:goal}
\end{eqnarray}
where
$g(\mc{N})=\prod_{r=0}^{m-1}\frac{{2m-2r\choose2}-\Psi_r(\mc{N})}{{2m-2r\choose2}-\psi_r}$
and the expectation $\mb{E}$ is with respect to the uniform distribution on the set of all
$m!$ orderings, $S(\mc{M})$. The goal is now to show that if
$G\in\mc{L}(\bar{d})$ then
\begin{equation}
\mb{E}(g(\mc{N}))\leq1+o(1).\label{eq:Eg<=1}
\end{equation}
Note that equations \eqref{eq:goal} and \eqref{eq:Eg<=1} finish the
proof. Thus, we only need to prove equation (\ref{eq:Eg<=1}).
\begin{proof}[of Equation (\ref{eq:Eg<=1})]
Before starting the proof it is important to see that
$g(\mc{N})=f(\mc{N})^{-1}$ and the aim of Section
\ref{sec:Kim-Vu-Calc} was to show that $\mb{E}(f(\mc{N}))=1+o(1)$. In
this section we will show that the concentration results of Section
\ref{sec:Kim-Vu-Calc} are strong enough to bound the variance of
$g(\mc{N})$ as well.

Recall the definitions for variables $\la_i$ and $T(\la_i)$ from Section
\ref{sec:Kim-Vu-Calc}. Here we will consider a different
partitioning of the set $S(\mc{M})$. Define subsets $F_0\subseteq
F_1\subseteq\ldots\subseteq F_L\subseteq S(\mc{M})$ as follows:
$$F_i=\{\mc{N}\in S(\mc{M})~|~\psi_r-\Psi_r(\mc{N})<T_r(\la_i):~\forall~0\leq r\leq m-\om\la_i/2\}$$
and $F_\infty=S(\mc{M})\stm\cup_{i=0}^LF_i$. The following two
lemmas are equivalent versions of Lemmas \ref{lem:f1A}, \ref{lem:f1C}.
\begin{lemma}\label{lem:gF} For all $1\leq i\leq L$,
\begin{itemize}
\item[(a)] $\mb{P}(F_i\stm F_{i-1})\leq e^{-\Om(\la_i)}$.
\item[(b)] For all $\mc{N}$ in $F_i\stm F_{i-1}$ we have $g(\mc{N})\leq e^{o(\la_i)}$.
\end{itemize}
\end{lemma}
\begin{lemma}\label{lem:gF0}
If $\mc{N}\in \mc{F}_0$ then  $g(\mc{N})\leq 1+o(1)$.
\end{lemma}
Proof of these Lemmas is similar to the proofs for Lemmas
\ref{lem:f1A} and \ref{lem:f1C}, and the only extra information that is required is
$$\sum_{2m-2r=2}^{\om\la}g(\mc{N})\leq2\f{\psi_r}{\f{(2m-2r)^2}{2}}=
O(\frac{\om\la d_{\max}^2}{m}).$$
Then for Lemma \ref{lem:gF} we use $\frac{\om\la
d_{\max}^2}{m}=o(\la)$ and for Lemma \ref{lem:gF0} we use
$\frac{\om\la_0 d_{\max}^2}{m}=o(1)$. The combination of these two lemmas gives $\mb{E}(g(\mc{N}))\leq 1+o(1)$.
\enp
\end{proof}
\subsection{Proof of Equation
(\ref{eq:varP=o(1)})}\label{subsec:varP=o(1)}

Similar to Section \ref{subsec:varN=o(1)} we will use lemmas from Section
\ref{sec:Kim-Vu-Calc}. The main technical point in this section
is a new result which exploits the combinatorial structure of the model to obtain a tighter bound than in Section \ref{sec:Kim-Vu-Calc}.

Equation (\ref{eq:varP=o(1)}) is equivalent to
$$\frac{\mb{E}_{\mrm{B}}(P^2)}{\mb{E}_{\mrm{B}}(P)^2}<1+o(1).$$
First notice that
$$
\frac{\mb{E}_{\mrm{B}}(P^2)}{\mb{E}_{\mrm{B}}(P)^2}=\frac{m!~\sum_{\mc{N}}\mb{P}_{\mrm{B}}(\mc{N})^2}{\mb{P}_{\mrm{B}}(G)^2}
=\frac{\mb{E}(f(\mc{N})^2)}{\mb{E}(f(\mc{N}))^2}.
$$
Therefore, all we need to show is $\mb{E}(f(\mc{N})^2)=1+o(1)$.

Consider the same partitioning of the set $S(\mc{M})$ as in
Section \ref{sec:Kim-Vu-Calc}. It is straightforward to see that
Lemmas \ref{lem:f1A}, \ref{lem:f1Ainf}, \ref{lem:f1B}, and \ref{lem:f1C}
give us the following stronger results as well
\begin{eqnarray*}
\mb{E}(f(\mc{N})^2\mrm{1}_{\mc{A}})&=&o(1),\\
\mb{E}(f(\mc{N})^2\mrm{1}_{\mc{B}})&=&o(1),\\
\mb{E}(f(\mc{N})^2\mrm{1}_{\mc{C}})&\leq&1+o(1).
\end{eqnarray*}
Thus, the only missing part is the following
\begin{equation}
\mb{E}(f(\mc{N})^2\mrm{1}_{S^*(\mc{M})\setminus
S^*(\mc{M})})=o(1)\label{eq:main-eq}
\end{equation}
which we will prove by using the combinatorial properties of the model.
\begin{proof}[of equation (\ref{eq:main-eq})]
Recall that $S^*(\mc{M})\setminus
S^*(\mc{M})$ consists of those orderings $\mc{N}$ that violate the condition
$$\Psi_r(\mc{N})\leq(1-\tau/4){2m-2r\choose2}\quad\quad\quad(\ast)$$
for some $r$. If this happens for some $r$ then from
Lemma \ref{lem:del-lam-bounds}(iii) and $d_{\max}^4=o(m)$ we have
\begin{multline*}
\Delta_r(\mc{N})\geq\Psi_r(\mc{N})-\f{d_{\max}^2}{8m}(2m-2r)^2\\
>\Psi_r(\mc{N})-\tau/4{2m-2r\choose2}>(1-\tau/2){2m-2r\choose2}.
\end{multline*}
On the other hand using Lemma \ref{lem:del-lam-bounds}(i) from
Section \ref{sec:Kim-Vu-Calc}: $\Delta_r(\mc{N})\leq
\f{d_{\max}^2(2m-2r)}{2}$. So for $2m-2r\geq\f{d_{\max}^2}{2-\tau}$ we have
$\Delta_r(\mc{N})\leq(1-\tau/2){2m-2r\choose2}$. Thus condition
$(*)$ is violated only for $r$ very close to $m$.  For these values of $r$ we use the following combinatorial lemma to find an upper bound for $f(\mc{N})$.
\begin{lemma}\label{lem:comb} For all $r$ such that $2m-2r\leq\f{d_{\max}^2}{2-\tau}$,
$$\f{{2m-2r\choose2}-\psi_r}{{2m-2r\choose2}-\Psi_r(\mc{N})}\leq 2d_{\max}.$$
%$$\Delta_r(\mc{N})\leq \f{(2m-2r)^2}{2}\left[1-\frac{2m-2r}{(d_{\max}+1)^2}\right]-(m-r)\leq %\left[1-\frac{2m-2r}{(d_{\max}+1)^2}\right]{2m-2r\choose2}$$
\end{lemma}
\begin{proof}  %When $(m-r)\leq d_{\max}$ the lemma follows from \eqref{eq:case(m-r)<d}.
Let $n_r$ be the number of available vertices ($v_i$'s with $W_i\neq 0$) at
step $r+1$.  Without loss of generality assume that all such vertices are $v_1,\ldots,v_{n_r}$.
For each $1\leq i\leq n_r$ let $\tilde{d}_i^{(r)}$ be the number of neighbors of $v_i$ among $v_1,\ldots,v_{n_r}$ at step $r+1$.  Then the number of suitable pairs at step $r+1$ is at least $1/2\sum_{i=1}^{n_r}(n_r-1-\tilde{d}_i^{(r)})d_i^{(r)}$. Now consider the cases $n_r\geq 2d_{\max}$ or $n_r< 2d_{\max}$ separately.
\begin{enumerate}
\item For $n_r\geq 2d_{\max}$ the number of suitable pairs at step $r+1$ is at least $1/2\sum_{i=1}^{n_r}(d_{\max})d_i^{(r)}=d_{\max}(m-r)$.  Therefore,
$$\f{{2m-2r\choose2}-\psi_r}{{2m-2r\choose2}-\Psi_r(\mc{N})}\leq \f{(m-r)(2m-2r-1)}{d_{\max}(m-r)(1-\f{d_{\max}^2}{4m})}\leq 2d_{\max} .$$
Here we used $d_{\max}^2=o(m)$ and $(2m-2r)\leq \f{d_{\max}^2}{2-\tau}\leq 3d_{\max}^2/5.$
\item For $n_r< 2d_{\max}$ we use $n_r\geq 1+\tilde{d}_i^{(r)}+d_{i}^{(r)}$ to show that the number of suitable pairs is at least $1/2\sum_{i=1}^{n_r}(n_r-1-\tilde{d}_i^{(r)})d_i^{(r)}\geq 1/2\sum_{i=1}^{n_r}(d_i^{(r)})^2\geq 1/2\f{(\sum_{i=1}^{n_r}d_i^{(r)})^2}{n_r}$.  Hence,
\begin{eqnarray}
\f{{2m-2r\choose2}-\psi_r}{{2m-2r\choose2}-\Psi_r(\mc{N})}&\leq& \f{(m-r)(2m-2r-1)}{\f{(m-r)(2m-2r)}{n_r}(1-\f{d_{\max}^2}{4m})}\nonumber\\
&\leq& n_r\f{1-\f{1}{d_{\max}^2}}{1-o(1)}\leq 2d_{\max}\nonumber.
\end{eqnarray}

\end{enumerate}
\enp\end{proof}
Lemma \ref{lem:comb} gives
$$
\prod_{r=m-t}^{m-1}\f{{2m-2r\choose2}-\psi_r}{{2m-2r\choose2}-\Psi_r(\mc{N})}\leq
2^td_{\max}^t.
$$
From here we will closely follow the steps taken in Section
\ref{subsec:pf-S*c-o1}. Since $t$ is the first place that $(*)$ is
violated
$$
\prod_{r=0}^{m-t-1}\f{{2m-2r\choose2}-\psi_r}{{2m-2r\choose2}-\Psi_r(\mc{N})}\leq
\exp\bigg[{\f{16}{\tau}\sum_{r=0}^{m-1}\f{\max(\Psi_r(\mc{N})-\psi_r,0)}{(2m-2r)^2}}\bigg].
$$
So,
\begin{eqnarray*}
f(\mc{N})\mrm{1}_{S_t}&=&\mrm{1}_{S_t}\prod_{r=0}^{m-1}\f{{2m-2r\choose2}-\psi_r}{{2m-2r\choose2}-\Psi_r(\mc{N})}\\
&\leq&
2^td_{\max}^t\mrm{1}_{S_t}\exp\bigg[{\f{16}{\tau}\sum_{r=0}^{m-1}\f{\max(\Psi_r(\mc{N})-\psi_r,0)}{(2m-2r)^2}}\bigg].
\end{eqnarray*}
Now using H\"{o}lder's inequality
\begin{multline*}
\mb{E}(f(\mc{N})^2\mrm{1}_{S_t})\\
\leq2^{2t}d_{\max}^{2t}\mb{E}\left(\mrm{1}_{S_t}\exp\bigg[{\f{32}{\tau}\sum_{r=0}^{m-t-1}\f{\max(\Psi_r(\mc{N})-\psi_r,0)}{(2m-2r)^2}}\bigg]\right)\\
\leq2^{2t}d_{\max}^{2t}\mb{E}(1_{S_t})^{1-\tau/2}\mb{E}\left(\mrm{1}_{S_t}\exp\bigg[{\f{64}{\tau}\sum_{r=0}^{m-t-1}\f{\max(\Psi_r(\mc{N})-\psi_r,0)}{(2m-2r)^2}}\bigg]\right)^{\tau/2}.
\end{multline*}
From Corollary \ref{cor:av-ep-gr} the second term
in the above product is $1+o(1)$ and we only need to show
$$
2^{2t}d_{\max}^{2t}\mb{P}(S_t)^{1-\tau/2}\leq\Big(1+o(1)\Big)\f{1}{m^{\tau
t}}.
$$
Now using the bound given by equation (\ref{eq:sum-d/m}) for
$\mb{P}(S_t)$ we have
\begin{eqnarray}
2^{2t}d_{\max}^{2t}2t^t\mb{P}(S_t)^{1-\tau/2}&\leq&\big(1+o(1)\big)2t\left(\f{2^{4-\tau/3}}{2-\tau}\f{d_{\max}^{4-5\tau+2\tau^2}}{m^{1-2.5\tau+\tau^2}}\right)^t\no\\
&\leq&\big(1+o(1)\big)2t\left(4\f{d_{\max}^{4-5\tau+2\tau^2}}{m^{1-2.5\tau+\tau^2}}\right)^t\no\\
&\leq&\big(1+o(1)\big)2t\left(4m^{-2.75\tau+3.5\tau^2-2\tau^3}\right)^t\no\\
&\leq&O\big(m^{-\tau t}\big).\no
\end{eqnarray}
\enp
\end{proof}
%---------------------------------------------------

\section{Acknowledgement}
We would like to thank Joe Blitzstein, Persi Diaconis, Adam Guetz,
Milena Mihail, Alistair Sinclair, Eric Vigoda and Ying Wang for insightful
discussions and useful comments on earlier version of this paper. We also thank the anonymous referees for their great comments and suggestions.
J.H. Kim was supported by the Korea Science and Engineering Foundation (KOSEF) grant funded by the Korea government(MOST) (No.
R16-2007-075-01000-0) and the second stage of the Brain Korea 21
Project in 2007. A. Saberi thanks the support of NSF.


\begin{thebibliography}{1}

%\bibitem{Chung2} W. Aiello, F. Chung and L. Lu, Random evolution in massive graphs,
%(2001), FOCS.
%
%\bibitem{ChungLuAiello} B. Aiello, F. Chung and L. Lu,
%A random graph model for massive graphs, (2000) STOC.
%

\bibitem{AS} N. Alon and J. Spencer, The Probabilistic Method, (1992), Wiley,
NY.

\bibitem{doyle} D. Alderson, J. Doyle, and W. Willinger, Toward and
Optimization-Driven Framework for Designing and Generating Realistic
Internet Topologies, (2002) HotNets.

%\bibitem{BarabasiAlbert} A. Barab\'{a}si and R. Albert,
%Emergence of scaling in random networks, , \emph{Science} (1999),
%286, 509-512.

%\bibitem{BayatiSaberi} M. Bayati, J.H. Kim and A. Saberi,
%Work in progress, (2006).

\bibitem{Montanari}
A. Amraoui, A. Montanari and R. Urbanke,  How to Find Good
Finite-Length Codes: From Art Towards Science, (2006) Preprint,
arxiv.org/pdf/cs.IT/0607064.

\bibitem{PersiBassetti} F. Bassetti, P. Diaconis,
Examples Comparing Importance Sampling and the Metropolis Algorithm,
(2005).

%\bibitem{BayatiSaberi} M. Bayati and A. Saberi,
%A sequential algorithm for generating random graphs, Preliminary
%version, (2006), http://www.stanford.edu/$\sim$bayati/rgperl.pdf.

\bibitem{BMS07}
M. Bayati, A. Montanari, and A. Saberi, Generating Random Graphs with Large Girth, (2009), \emph{ACM-SIAM Symposium on Discrete Algorithms (SODA)}.

%\bibitem{BKS06}
%M. Bayati, J. H. Kim and A. Saberi, A Sequential Algorithm for
%Generating Random Graphs, (2006), preprint.

\bibitem{BKS07}
M. Bayati, J. H. Kim and A. Saberi, A Sequential Algorithm for
Generating Random Graphs, (2007), International Workshop on Randomization and Computation.

\bibitem{BenderCanfield} E.A. Bender and E.R. Canfield,
The asymptotic number of labeled graphs with given degree sequence,
\emph{J. Combinatorial Theory Ser.} (1978), A 24, 3, 296-307.

\bibitem{BezakovaBhatnagarVigoda} I. Bez\'{a}kov\'{a}, N. Bhatnagar and E. Vigoda
Sampling Binary Contingency tables with a Greedy Start, (2006),
SODA.

\bibitem{BezakovaSinclairStephanovicVigoda} I. Bez\'{a}kov\'{a}, A. Sinclair, D. \u{S}tefankovi\v{c} and E. Vigoda
Negative Examples for Sequential Importance Sampling of Binary
Contingency Tables, Proc. of Annual European Symp., Vol (14), (2006).

\bibitem{blanchet} J. Blanchet, Eficient Importance Sampling for Binary Contingency Tables, Ann. Appl. Probability, Vol (19), No. 3, pp 949-982, (2009).

\bibitem{JoePersi} J. Blitzstein and P. Diaconis, A sequential
importance sampling algorithm for generating random graphs with
prescribed degrees, (2005), Submitted.

\bibitem{Bollobas1980} B. Bollob\'{a}s, A probabilistic proof of an
asymptotoic forumula for the number of labelled regular graphs,
(1980), \emph{European J. Combin.} 1, 4, 311-316.

%\bibitem{BollobasBook} B. Bollob\'{a}s, Random Graphs, (2001), Cambrdige University Press.

%\bibitem{BrittonDeijfenLof} T. Britton, M. Deijfen and A. Martin-L\"{o}f ,Generating simple random graphs with
%prescribed degree distribution, (2005), Preprint.

\bibitem{bu} T. Bu and D. Towsley, On Distinguishing between
Internet Power Law Topology Generator, (2002) INFOCOM.

%\bibitem{chen} Q. Chen, H. Chang, R. Govindan, S> Jamin, S. Shenker
%and W Willinger, The Origins of Power-Laws in Internet Topologies
%Revisited, (2002) Infocom.

\bibitem{ChenDiaconisHolmsLiu} Y. Chen, P. Diaconis, S. Holmes and
J.S. Liu, Sequential Monte Carlo methods for statistical analysis of
tables, (2005), \emph{Journal of the American Statistical
Association} 100, 109-120.

\bibitem{CooperDyerGreenhill} C. Cooper, M. Dyer and C. Greenhill, Sampling regular graphs and peer-to-peer network, (2007), \emph{Combinatorics, Probability and Computing }, vol 16.

\bibitem{ChungLu} F. Chung and L. Lu, Conneted components in random graphs with given expected degree sequence, (2002), \emph{Ann. Comb. } 6, 2, 125-145.

\bibitem{DiaconisGangolli} P. Diaconis and A. Gangolli, Rectangular arrays with fixed margins, (1995), \emph{Discrete probability and algorithms (Minneapolis, MN, 1993) }. IMA Vol. Math. Appl., Vol 72. Springer, New York, 15-41.

%\bibitem{DurretBook} R. Durret, Random Graph Dynamics, (2006),
%Cambridge University Press.

\bibitem{faloutsos} M. Faloutsos, P. Faloutsos and C. Faloutsos, On
Power-law Relationships of the Internet Topology, (1999) SIGCOM.

%\bibitem{FederSaberi} T. Feder and A. Guetz and M. Mihail and A. Saberi, The flip markov chain and sampling connected graphs (2005), submitted.

\bibitem{GkantsidisMihailZegura}
C. Gkantsidis, M. Mihail and E. Zegura, The Markov Chain Simulation
Method for Generating Connected Power Law Random Graphs, (2003)
Alenex.


\bibitem{JerrumValiantVazirani} M. Jerrum, L. Valiant and V.
Vazirani, Random generation of combinatorial structures from a
uniform distribution, (1986), \emph{Theoret. Comput. Sci.} 43,
169-188. 73, 1, 91-100.

\bibitem{JerrumSinclair1989} M. Jerrum and A. Sinclair, Approximate counting, uniform generation and rapidly mixing Markov chains,
(1989), \emph{Inform. and Comput.} 82  ,  no. 1, 93--133.


\bibitem{JerrumSinclair1990} M. Jerrum and A. Sinclair, Fast uniform
generation of regular graphs, (1990), \emph{Theoret. Comput. Sci.}
73, 1, 91-100.

\bibitem{JerrumSinclairMcKay} M. Jerrum, A. Sinclair and B. McKay,
When is a graphical sequence stable?, (1992) \emph{Random graphs Vol
2 (Pozna\'{n} 1989)} Wiley-Intersci. Publ. Wiley, New York, 101-115.

%\bibitem{JerrumSinclairVigoda} M. Jerrum, A. Sinclair and E. Vigoda,
%A Polynomial-Time Approximation Algorithm for the Permanent of a
%Matrix with Non-Negative Entries, (2004) \emph{Journal of the ACM},
%51(4):671-697.

%\bibitem{jin}
%C. Jin, Q. Chen and S. Jamin, Inet: Internet Topology Generator,
%University of Michigan Technical Report, CSE-TR-433-00. Available at
%http://irl.eecs.umich.edu/jamin.

\bibitem{KannanTetaliVempala} R. Kannan, P. Tetali and S. Vempala,
Simple Markov chain algorithms for generating bipartite graphs and
tournaments, (1992) \emph{Random Structures and Algorithms} (1999)
14, 293-308.


\bibitem{Kim} J.H.Kim , On Brooks' Theorem for Sparse Graphs, \emph{Combi. Prob. \& Comp.}, (1995)  4, 97-132.

\bibitem{Kim-Vu2000} J. H. Kim and V. H. Vu, Concentration of multivariate polynomials and its applications,
(2000) \emph{Combinatorica} 20, no 3, 417-434.

\bibitem{Kim-Vu} J. H. Kim and V. H. Vu, Generating Random Regular Graphs,
, STOC 2003 213-222.

\bibitem{KimVuSandwitch} J.H.Kim and V. Vu, Sandwiching random graphs, \emph{Advances in Mathematics} (2004)
188, 444-469.

\bibitem{knuth} D. Knuth, Mathematics and computer science: coping with finiteness, (1976) \emph{Science} 194(4271):1235-1242.

\bibitem{medina}
A. Medina, I. Matta and J. Byers, On the origin of power laws in
Internet topologies, \emph{ACM Computer Communication Review},
(2000), vol. 30, no. 2, pp. 18-28.

\bibitem{Mckay} B. McKay , Asymptotics
for symmetric 0-1 matrices with prescribed row sums, (1985),
\emph{Ars Combinatorica} 19A:15-25.

%\bibitem{MckayWormald1990a} B. McKay and N. C. Wormald, Asymptotic
%enumeration by degree sequence of graphs of high degree, (1990a),
%\emph{European J. Combin.} 11, 6, 565-580.

\bibitem{MckayWormald1990b} B. McKay and N. C. Wormald, Uniform generation of
random regular graphs of moderate degree, (1990b), \emph{J.
Algorithms} 11, 1, 52-67.

\bibitem{MckayWormald1991} B. McKay and N. C. Wormald, Asymptotic
enumeration by degree sequence of graphs with degrees o($n^{1/2}$),
(1991), \emph{Combinatorica} 11, 4, 369-382.


\bibitem{MiloKashtanItzkovitzNewmanAlon} R. Milo, N. Kashtan, S. Itzkovitz, M. Newman and U. Alon, On the uniform
generation of random graphs with prescribed degree sequences,
(2004), http://arxiv.org/PS\_cache/cond-mat/pdf/0312/0312028.pdf

\bibitem{MSOIKCA} R. Milo,  S. ShenOrr, S. Itzkovitz, N. Kashtan, D. Chklovskii and U.
Alon, Network motifs: Simple building blocks of complex networks,
(2002), \emph{Science} 298, 824-827


\bibitem{MolloyReed1995} M. Molloy and B. Reed, A critical point for random graphs with a given degree sequence,
(1995), \emph{Random Structures and Algorithms} 6, 2-3, 161-179.

%\bibitem{MolloyReed1998} M. Molloy and B. Reed, The size of the ginat component of a random graph with a given degree sequence,
%(1998), \emph{Combin. robab. Comput.} 7, 3, 295-305.

%\bibitem{Newman2003}
%M. Newman, The structure and function of complex networks, (2003),
%\emph{SIAM review} 45, 167-256.

%\bibitem{NewmanStrogatzWatts2001}
%M. Newman, S. Strogatz and D. Watts, Random graphs with arbitrary
%degree distributions and their applications, (2001), \emph{Physical
%Review} E 64, 026118.
\bibitem{AlistairEmail} A. Sinclair, Personal communication, (2006).

%\bibitem{palmer} C. Palmer and J. Steffan, Generating network
%topologies that obey power laws, (2000) Globecom.


\bibitem{StegerWormald} A. Steger and N.C. Wormald, Generating random regular graphs quickly,
(English Summary) Random graphs and combinatorial structures
(Oberwolfach, 1997), \emph{Combin. Probab. Comput.} 8, no. 4,
377-396.

%\bibitem{Tinhofer1979} G. Tinhofer, On the generation of random
%graphs with given properties and known distributions, (1979),
%\emph{Applied Computer Science Berichte Praktische Informatik} 13,
%265-296.

%\bibitem{Tinhofer1990} G. Tinhofer, Generating graphs uniformly at random, (1990),
%\emph{Computing Supplementum} 7, 235-255.

\bibitem{inet} H.Tangmunarunkit, R.Govindan, S.Jamin, S.Shenker, and W.Willinger,
Network Topology Generators: Degree based vs. Structural, (2002),
ACM SIGCOM.

%\bibitem{inet2}  http://topology.eecs.umich.edu/inet/

%\bibitem{Wormald1984} N. C. Wormald, Generating random regular
%graphs, (1984), \emph{J. Algorithms} 5, 2, 247-280.

\bibitem{Wormald1999} N. C. Wormald, Models of random regular
graphs, (1999), \emph{Surveys in combinatorics (Canterbury)} London
Math. Soc. Lecture Notes Ser., Vol 265. Cambridge Univ. Press,
Cambridge, 239-298.

%\bibitem{Valiant1979} L. G. Valiant, The complexity of computing the permanent,
%\emph{Theoretical Computer Science} 8 (1979), 189-201.

\bibitem{Vu} V. H. Vu, Concentration of non-Lipschitz functions and applications,
Probabilistic methods in combinatorial optimization, \emph{Random
Structures Algorithms} 20 (2002), no. 3, 267-316.

%\bibitem{zegura}, E. Zegura, K. Calvert, M. Donahoo, How to model
%and Internetwork, (1996) Infocom.
\end{thebibliography}
\end{document}